# Handover Management in Highly Dense Femtocellular Networks

Mostafa Zaman Chowdhury and Yeong Min Jang*

*Department of Electronics Engineering, Kookmin University, Seoul 136-702, Korea.*

E-mail: mzceee@yahoo.com, yjang@kookmin.ac.kr

∗Corresponding author

**Abstract** Femtocell technology is envisioned to be widely deployed in subscribers' homes to provide high data-rate communications with quality of service. Dense deployment of femtocells will offload large amounts of traffic from the macrocellular network to the femtocellular network by the successful integration of macrocellular and femtocellular networks. Efficient handling of handover calls is the key for successful femtocell/macrocell integration. For dense femtocells, intelligent integrated femtocell/macrocell network architecture, a neighbor cell list with a minimum number of femtocells, effective call admission control (CAC), and handover processes with proper signaling are the open research issues. An appropriate traffic model for the integrated femtocell/macrocell network is also not yet developed. In this paper, we present the major issue of mobility management for the integrated femtocell/macrocell network. We propose a novel algorithm to create a neighbor cell list with a minimum, but appropriate, number of cells for handover. We also propose detailed handover procedures and a novel traffic model for the integrated femtocell/macrocell network. The proposed CAC effectively handles various calls. The numerical and simulation results show the importance of the integrated femtocell/macrocell network and the performance improvement of the proposed schemes. Our proposed schemes for dense femtocells will be very effective for those in research and industry to implement.

**Keywords** *Femtocell, dense femtocell, handover, SON, neighbor cell list, femtocell-to-femtocell handover, macrocell-to-femtocell handover, femtocell-to-macrocell handover, traffic model, and CAC.*

## 1. Introduction

Future wireless networks will necessitate high data-rates with improved quality of service (QoS) and low cost. A femtocellular network [1-9] is one of the most promising technologies to meet the tremendous demand of increasing wireless capacity by various wireless applications for future wireless communications. Femtocells operate in the spectrum licensed for cellular service providers. The key feature of the femtocell technology is that users require no new equipment (UE). The deployment cost of the femtocell is very low while providing a high data rate. Thus, the deployment of femtocells at a large scale [5, 6] is the ultimate objective of this technology. Indeed, a well-designed femtocell/macrocell integrated network can divert huge amounts of traffic from congested and expensive macrocellular networks to femtocellular networks. From the wireless operator point of view, the ability to offload a large amount of traffic from macrocellular networks to femtocellular networks is the most important advantage of the femtocell/macrocell integrated network architecture. This will not only reduce the investment capital, the maintenance expenses, and the operational costs but will also improve the reliability of the cellular networks [5].

Fig. 1 shows an example of femtocellular network deployment. The femtocells are deployed under the macrocellular network coverage or in a separate non-macrocellular coverage area. In the overlaid macrocell coverage area, femtocell-to-femtocell, femtocell-to-macrocell, and macrocell-to-femtocell handovers occur owing to the deployment of femtocells. The frequency of these handovers increases as the density of femtocells is increased. Thus, effective handover mechanisms are essential to support these handovers. The efficient femtocell-to-femtocell and femtocell-to-macrocell handovers result in seamless movement of femtocell users. Even though the macrocell-to-femtocell handover is not essential for seamless movement, efficient handling of this handover type can reduce huge traffic loads of macrocellular networks by transferring the calls to femtocells.

The large- and dense-scale deployment of femtocells suffers from several challenges [2-5]. Handover is one challenging issue among several issues. For efficient handover management, four factors, namely, intelligent network support, signal flow control for the handovers, reduced neighbor cell list, and an effective call admission control (CAC) policy are essential. To the best of our knowledge, complete research results regarding these issues are still unpublished. However, a few research groups (e.g., [10, 11]) have partially discussed some ideas regarding handover issues in femtocellular networks. T. Bai *et. al.* [10] proposed a handover mechanism based on the decision made by an entity connected with a femtocell access point (FAP). This entity considers the user type, access mode of the FAP, and current load of the FAP to make a decision about the target femtocell. However, their scheme does not consider the creation of a neighbor cell list. H. Zhang *et. al.* [11] presented a handover optimization algorithm based on the UE's mobility state. They also presented an analytical model for the handover signaling cost analysis. Here, we propose some novel approaches to solve the mobility management issues for densely deployed femtocellular networks. We suggest self-organizing network (SON) features to support the dense femtocellular networks, detail handover call flows for different handovers, an algorithm to create an appropriate neighbor cell list (including the neighbor femtocell list and the neighbor macrocell list), and an efficient CAC to handle various calls. We also propose a novel traffic model for the integrated femtocell/macrocell scenario.

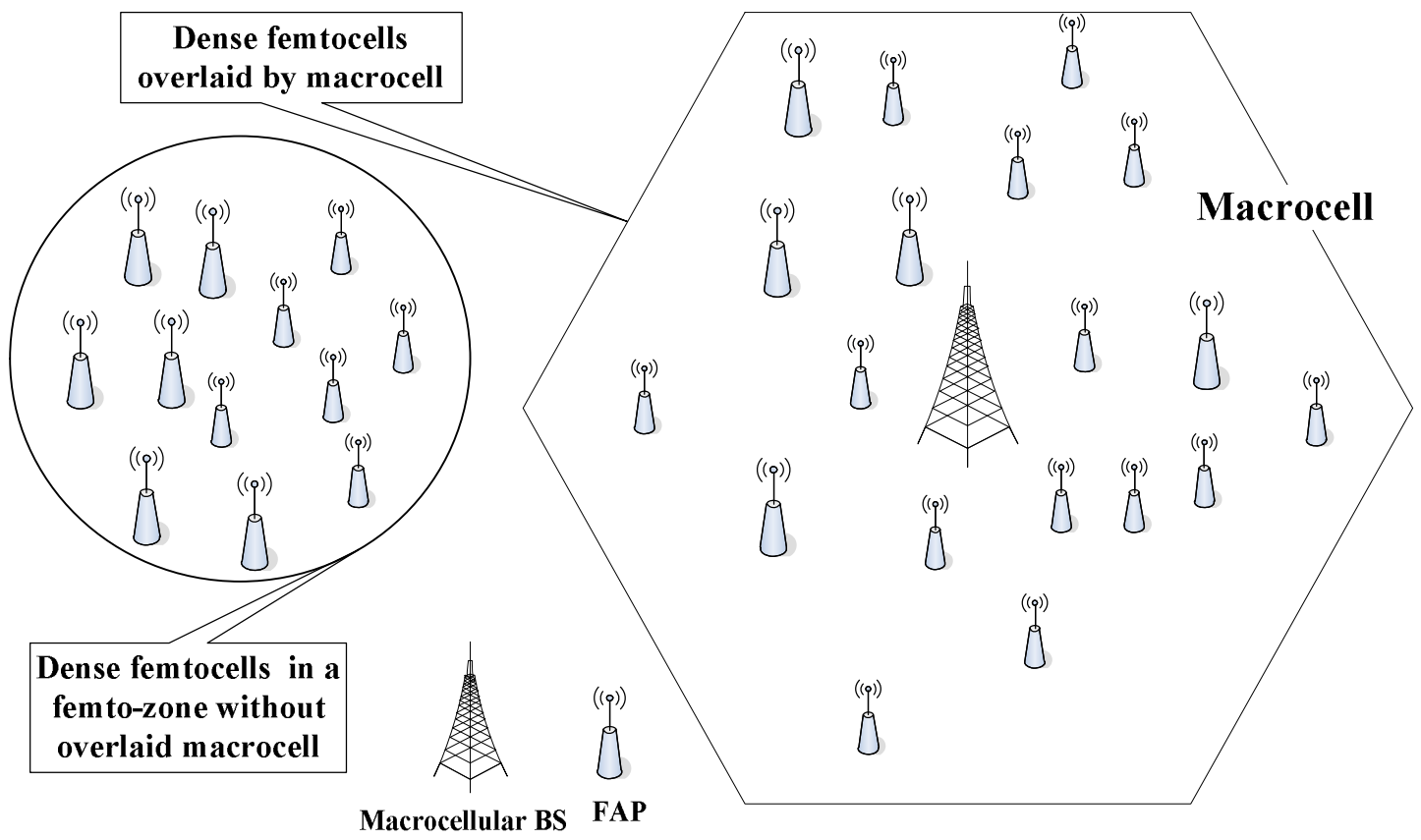

**Fig. 1.** Example of a dense femtocellular network deployment scenario.

When the number of femtocells increases, the system architectures must support the efficient management of a large number of FAPs and a huge number of handover calls. The SON features [5, 12, 13] can support the coordination among the FAPs as well as among the FAPs and macrocellular BS to execute smooth handover.

The ability to seamlessly move between the macrocellular network and the femtocellular networks is a key driver for femtocell network deployment. Moreover, handover between two networks should be performed with minimum signaling. Owing to some modifications of the existing network and protocol architecture for integrated femtocell/macrocell networks, the proposed signal flows for handover procedures are slightly different as compared to the macrocellular case.

In a dense femtocellular network deployment, thousands of femtocells can be deployed within a small coverage area. As a result, this may present huge interference effects. Whenever a mobile station (MS) realizes that the received signal from the serving FAP is going down, the MS may receive multiple signals from several of the neighbor FAPs for handover. Thus, the neighbor cell list based on the received signal only will contain a large number of femtocells. In addition, a hidden FAP problem may arise. The hidden FAP problem arises when a neighbor FAP is very close to the MS but the MS cannot receive the signal owing to some barrier (e.g., a wall) between the MS and that FAP. Thus, the hidden FAPs will be out of the neighbor cell list if the neighbor femtocell list is designed on the basis of the received signals only. The same incidences are also applicable for the macrocell-to-femtocell handover case. The proposed algorithms are capable of providing a neighbor cell list that contains a minimum number of femtocells as well as includes the hidden FAPs.

The proposed CAC does not differentiate between the new originating calls and handover calls for the femtocellular networks owing to available resources in the femtocellular networks. The CAC provides higher priority for the handover calls in the overlaid macrocellular network by offering a QoS adaptation provision [14, 15]. The QoS adaptation provision is only available to accept handover calls in a macrocellular network. Thus, the macrocellular network can accept a large number of handover calls that are generated because of the femtocells and the neighbor macrocells. The CAC policy also offers two levels of signal-to-noise plus interference ratio (SNIR) thresholds to reduce some unnecessary macrocell-to-femtocell handovers.

The existing traffic model should be modified such that it can be applied to integrated networks. We propose a novel traffic model for femtocell/macrocell integrated networks that is useful to analyze the performance of femtocell/macrocell integrated networks.

The rest of this paper is organized as follows. Section 2 suggests the system network architecture to support dense femtocells. The SON features of the network architecture are also proposed in this section. The neighbor cell list management algorithms are proposed in Section 3. In Section 4, we describe the call flows for the macrocell-to-femtocell, femtocell-to-femtocell, and femtocell-to-macrocell handovers. CAC policies are provided in Section 5. In Section 6, we derive the detailed traffic model and queuing analysis for the femtocell/macrocell integrated networks. Performance evaluation results of the proposed schemes are presented and compared in Section 7. Finally, Section 8 concludes our work.

## 2. Network Architecture to Support Dense Femtocells

In this section, we discuss the network architecture to support dense femtocells. Fig. 2 shows one example of concentrator-based device-to-core network (CN) connectivity for femtocell/macrocell integrated networks to support dense femtocellular networks [1, 2, 4-7, 16, 17]. Several FAPs are connected to a femto gateway (FGW) through a broadband ISP or another network. The FGW acts like a concentrator and also provides Security Gateway functionalities for the connected FAPs. The FGW communicates with the RNC through the CN. There is no direct interface between the RNC and the FGW. The FGW entity appears as a legacy RNC to the existing CN. The FGW manages the traffic flows for thousands of femtocells. Traffic from different access networks comes to the FGW and is then sent to the desired destination networks. There is interoperability between the femtocell operator and the ISP network or other mobile operators to connect the femtocell users with other users from that operator. The service level agreement (SLA) between the femtocell operator and the ISP network operator ensures sufficient bandwidth for the femtocell users. Whenever an FAP is installed, the respective FGW provides the FAP’s position and its authorized user list to the macrocellular BS database (DB) server through the CN.

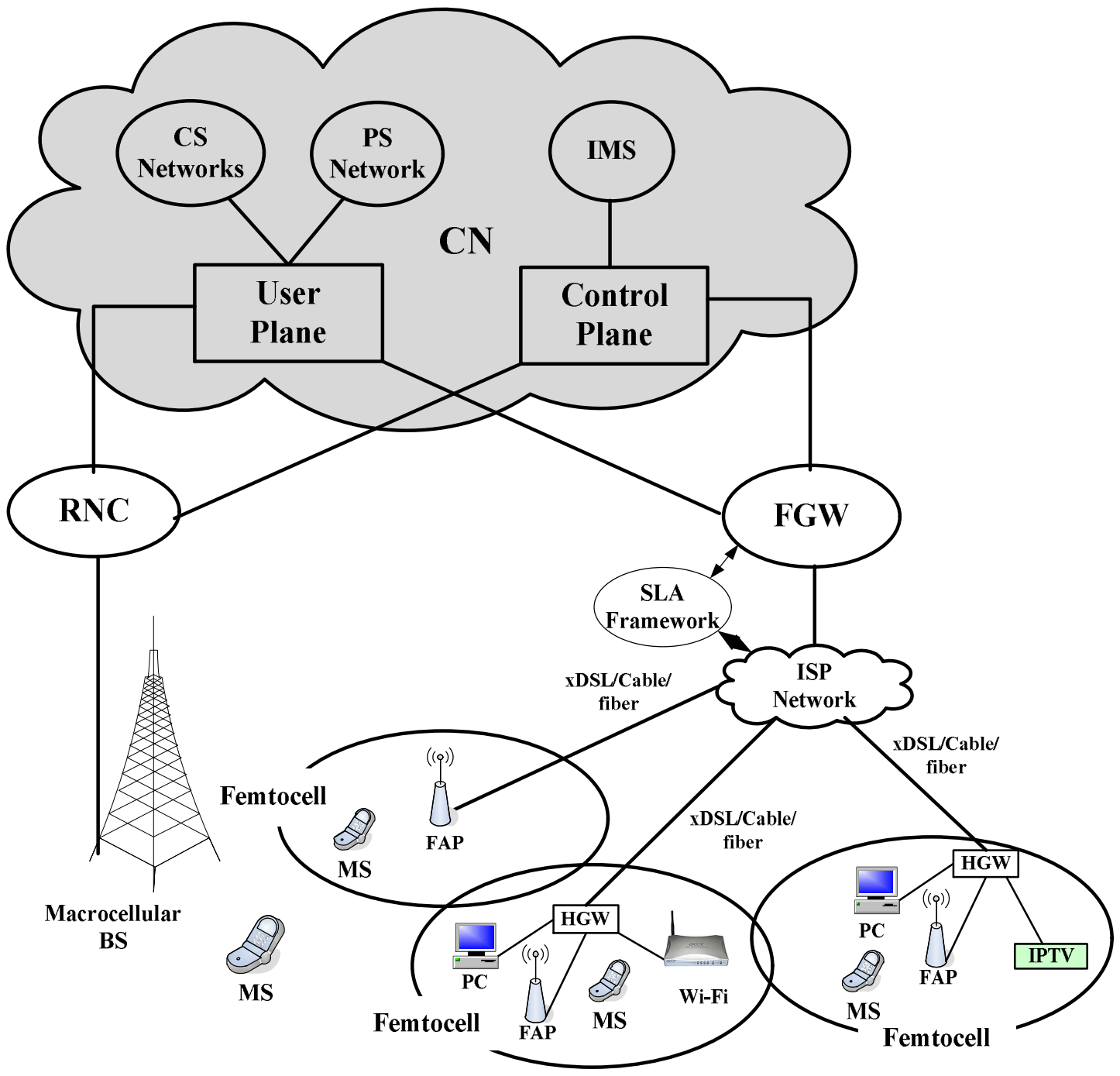

**Fig. 2.** Example of device-to-CN connectivity for dense femtocellular network deployment.

From the network operator's perspective, the main requirement for dense femtocell deployment is that it fits into the network with minimum level of operator involvement in the deployment process while minimizing the impact of the femtocell on the existing network. For this purpose, the femtocell is required to boot up into a network by sniffing so that it can scan the air interface for available frequencies and other network resources. Self-organization of radio access networks is regarded as a new approach that enables cost-effective support of a range of high-quality mobile communication services and applications for acceptable prices. It enables deployment of dense femtocell clusters, providing advanced SON mechanisms [6, 12, 13] generally eliminating interference between femtocells, as well as reducing the size of the neighbor cell list and scanning for the handover to ensure fast and reliable handover.

The main functionalities of the SON for femtocellular networks are self-configuration, self-optimization, and self-healing [6, 13]. Self-configuration includes frequency allocation. Self-optimization includes transmission power optimization, neighbor cell list optimization, coverage optimization, and mobility robustness optimization. Self-healing includes automatic detection and solution of most of the failures. Neighbor FAPs as well as the macrocellular BS and the neighbor FAPs coordinate with each other. Whenever an MS desires handover in an overlaid macrocell environment, the MS detects multiple neighbor FAPs because of the dense deployment of femtocells along with the presence of macrocell coverage. Thus, during the handover phase, it is quite difficult to sense the actual FAP to which the user is going to be handed over to. The location information is exchanged among the neighbor FAPs as well as among the neighbor FAPs and macrocellular BS for building an optimized neighbor femtocell list. The handover processes are facilitated by the SON features of the network.

## 3. Neighbor Femtocell List

Finding the neighbor FAPs and determining the appropriate FAP for the handover are challenges for optimum handover decision [5]. Macrocell-to-femtocell and femtocell-to-femtocell handovers in a dense femtocellular network environment suffer from some additional challenges because of dense neighbor femtocells. In these handovers, the MS needs to select the appropriate target FAP among many neighbor FAPs. These handovers create significant problems if there is no minimum number of femtocells in the neighbor femtocell list. The MSs use much more power consumption in order to scan multiple FAPs, and the MAC overhead becomes significant. This increased size of the neighbor femtocell list along with messaging and broadcasting a large amount of information causes too much overhead. Therefore, an appropriate and optimal neighbor femtocell list is essential for dense femtocellular network deployment.

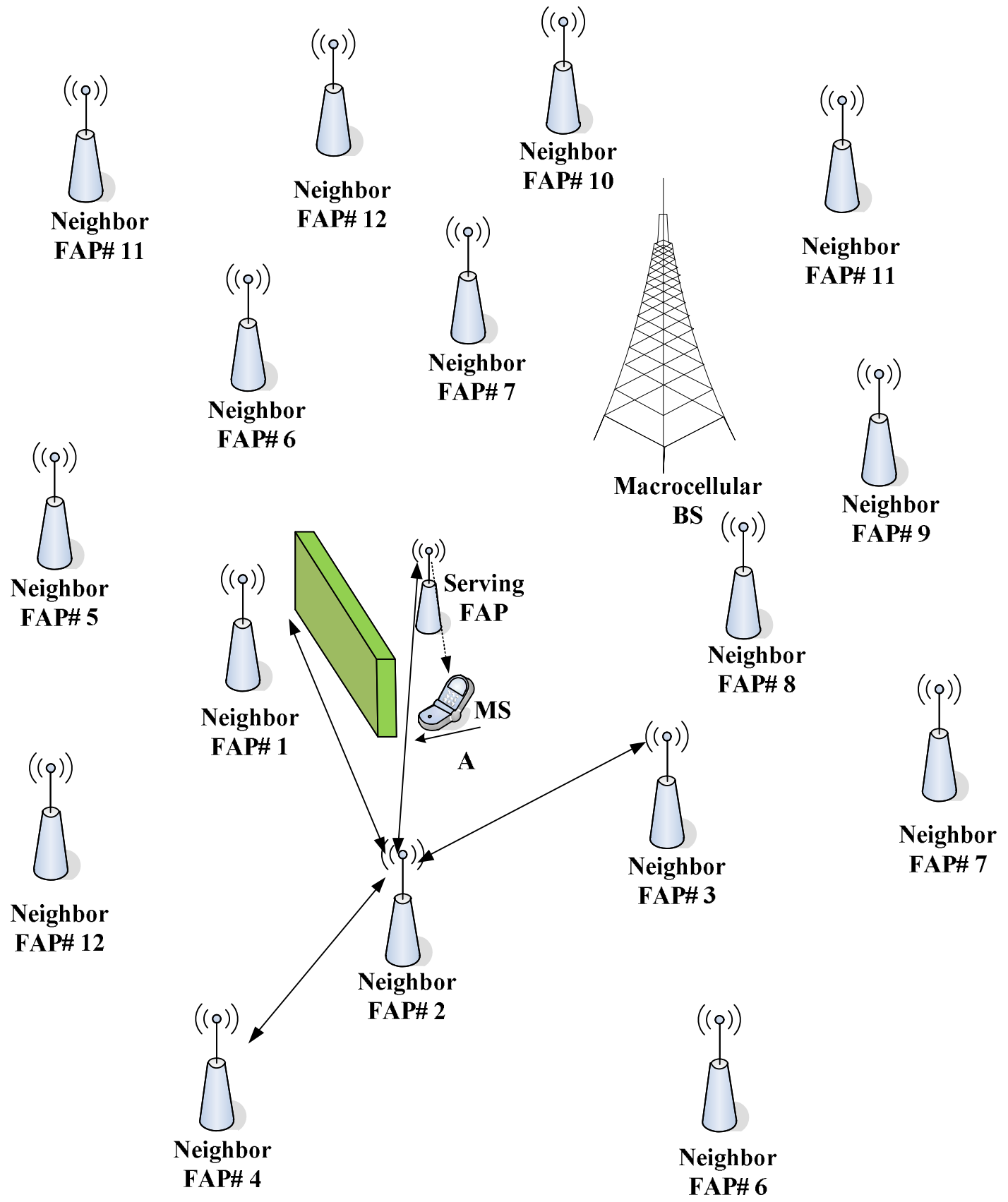

**Fig. 3.** Scenario of dense femtocellular network deployment where several hidden FAPs and other FAPs are situated as neighbor femtocells.

Whenever an MS moves away from one femtocell or the MS moves around the macrocellular coverage area, the MS detects signals from many neighbor FAPs owing to dense deployment of femtocells while detecting the presence of macrocell coverage. Reducing the size of the neighbor femtocell list is essential to minimize the amount of scanning and signal flow during handover. A large neighbor femtocell list causes unnecessary scanning for the handover. Traditional schemes (e.g., [18, 19]) based on the received signal strength indicator (RSSI) are used for the existing cellular system. However, the neighbor femtocell list based on only the RSSI will contain a large number of femtocells in the list. Therefore, these traditional schemes are not effective for creating the neighbor femtocell list in a dense femtocellular network environment. In addition, missing some of the hidden femtocells in the neighbor femtocell list causes the failure of handover. Our main objective is to create such a neighbor femtocell list for the femtocell-to-femtocell and macrocell-to-femtocell handovers so that the list contains the minimum number of femtocells and considers all the hidden femtocells. The FAPs and the macrocellular BS coordinate with each other to facilitate a smooth handover in our proposed scheme. Fig. 3 shows a scenario of dense femtocellular network deployment where several FAPs are situated as neighbor femtocells. For the MS at position “A,” the MS cannot receive a sufficient signal level from FAP# 1 because of a wall or another obstacle between the MS and this FAP. The serving FAP and FAP# 1 also cannot coordinate with each other. Thus, a neighbor femtocell list based on the RSSI measurement does not include FAP# 1 in the neighbor femtocell list. In this situation, FAP# 2 and FAP# 1 coordinate with each other using the SON features. FAP# 2 gives the location information of FAP# 1 to the serving FAP. Once receiving this location information, the neighbor femtocell list includes FAP# 1. Therefore, the MS can complete the pre-handover processes with FAP# 1, with coordination between the serving FAP and FAP# 1, even though the MS cannot receive the signal from FAP# 1. Subsequently, if the MS moves closer to FAP# 1, receives a sufficient level of signal from FAP# 1, and the received signal from the serving FAP goes below the threshold level then connection is handed over from the serving FAP to FAP# 1.

Figs. 4 and 5 show the flow mechanisms for the design of the optimal neighbor femtocell list. $N_f$ and $N_c$ denote the total number of femtocells and cells included in the neighbor cell list, respectively. Our proposed scheme initially considers the received RSSI level to create the neighbor cell list. For dense femtocellular network deployment, the frequency for each of the FAPs is allocated on the basis of the neighboring overlapping femtocells. Thus, the overlapping of the two femtocells does not use the same frequency to avoid interference [6]. The same frequency is only used by femtocells located far enough apart. Therefore, for the femtocell-to-femtocell handover case, the FAPs are removed from the initial neighbor femtocell list on the basis of the RSSI level of only those that use the same frequency as the serving FAP. Finally, hidden femtocells in the neighbor femtocell list are added using the location information coordination among neighbor FAPs or among the neighbor FAPs and macrocellular BS.

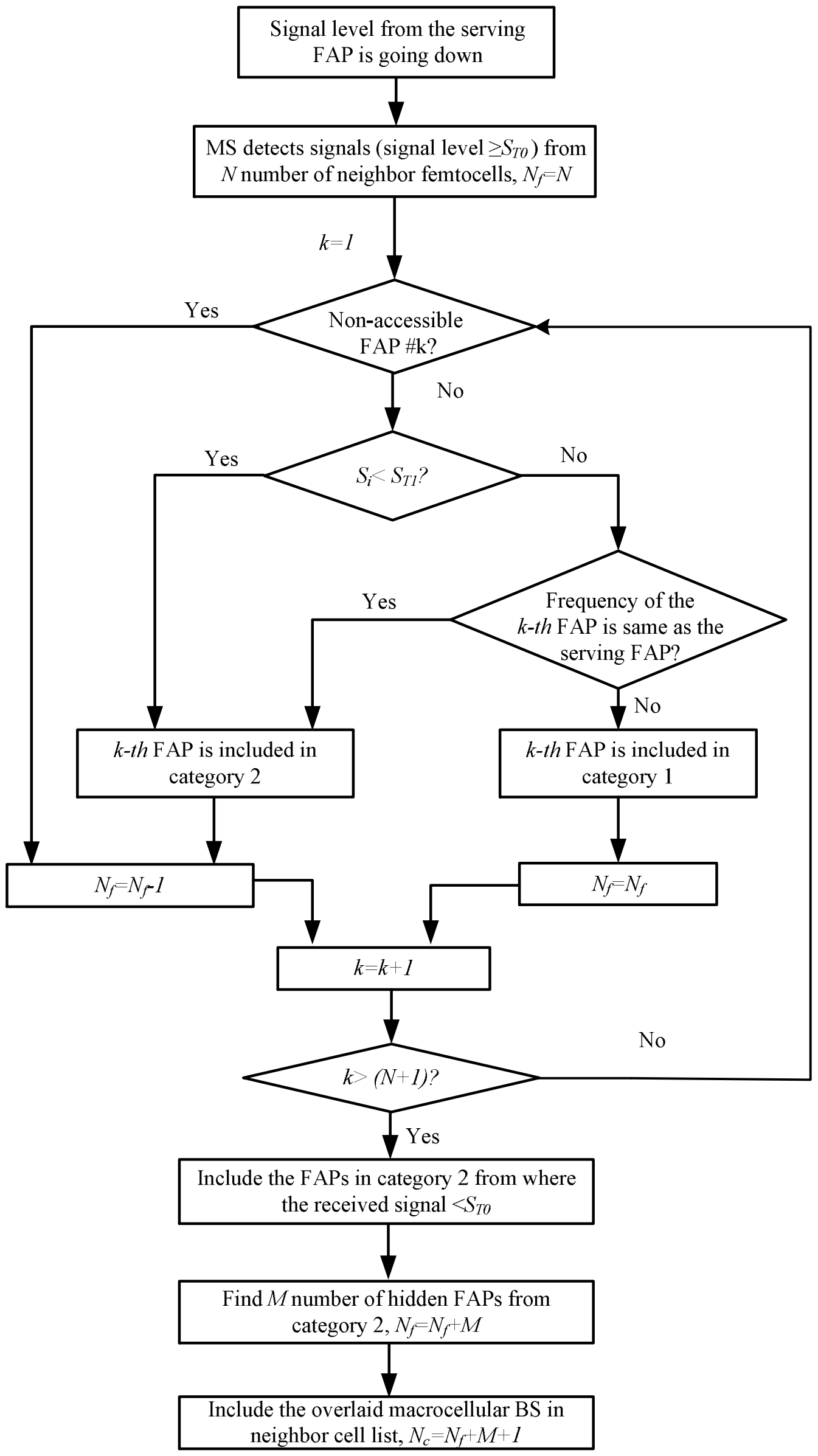

**Fig. 4.** Flow mechanism for the design of the optimal neighbor cell list for handover when the MS is connected with an FAP.

Fig. 4 describes the flow mechanism for the design of the optimal neighbor cell list for the handover when the MS is connected with an FAP. Fig. 5 describes the flow mechanism for the design of the optimal neighbor cell list for the handover when the MS is connected with the overlaid macrocellular network. We use two threshold levels of a signal to design the flow mechanisms. The first threshold signal level $S_{T0}$ is the minimum level of RSSI that is required to detect the presence of an FAP. The second signal level $S_{T1}$ is higher than $S_{T0}$. This level of RSSI is considered in our proposed scheme to build up the neighbor cell list. The criterion used for determining the value of $S_{T1}$ is the density of femtocells. Therefore, by increasing the value of $S_{T1}$ with the increasing density of femtocells, the number of femtocells in the neighbor cell list can be reduced. This action also reduces unnecessary handovers and the ping-pong effect. After checking the open/closed access [20] system, the *k-th* FAP is directly added to the neighbor cell list if the received signal $S_i$ from the *k-th* FAP is greater than or equal to the second threshold $S_{T1}$. All $N$ number of FAPs from where the MS receives signals are initially considered to create the neighbor cell list. Then, for the closed access case, all the non-accessible FAPs are removed from the number of initially considered femtocells. The frequency allocations are considered to find out the nearest FAPs for possible handover. The coordination among the neighbor FAPs as well as among the FAPs and macrocellular BS are performed to find hidden FAPs. Hidden FAPs are those from which the received signals are less than the second signal level $S_{T1}$; however, these FAPs are very close to the serving FAP. Even though these FAPs are very close to the MS, it receives a low level of signal or no signal from these FAPs owing to some obstacle between the MS and these FAPs. Thus, the addition of these hidden FAPs in the neighbor cell list reduces the chance that the MS fails to perfectly handover to the target FAP.

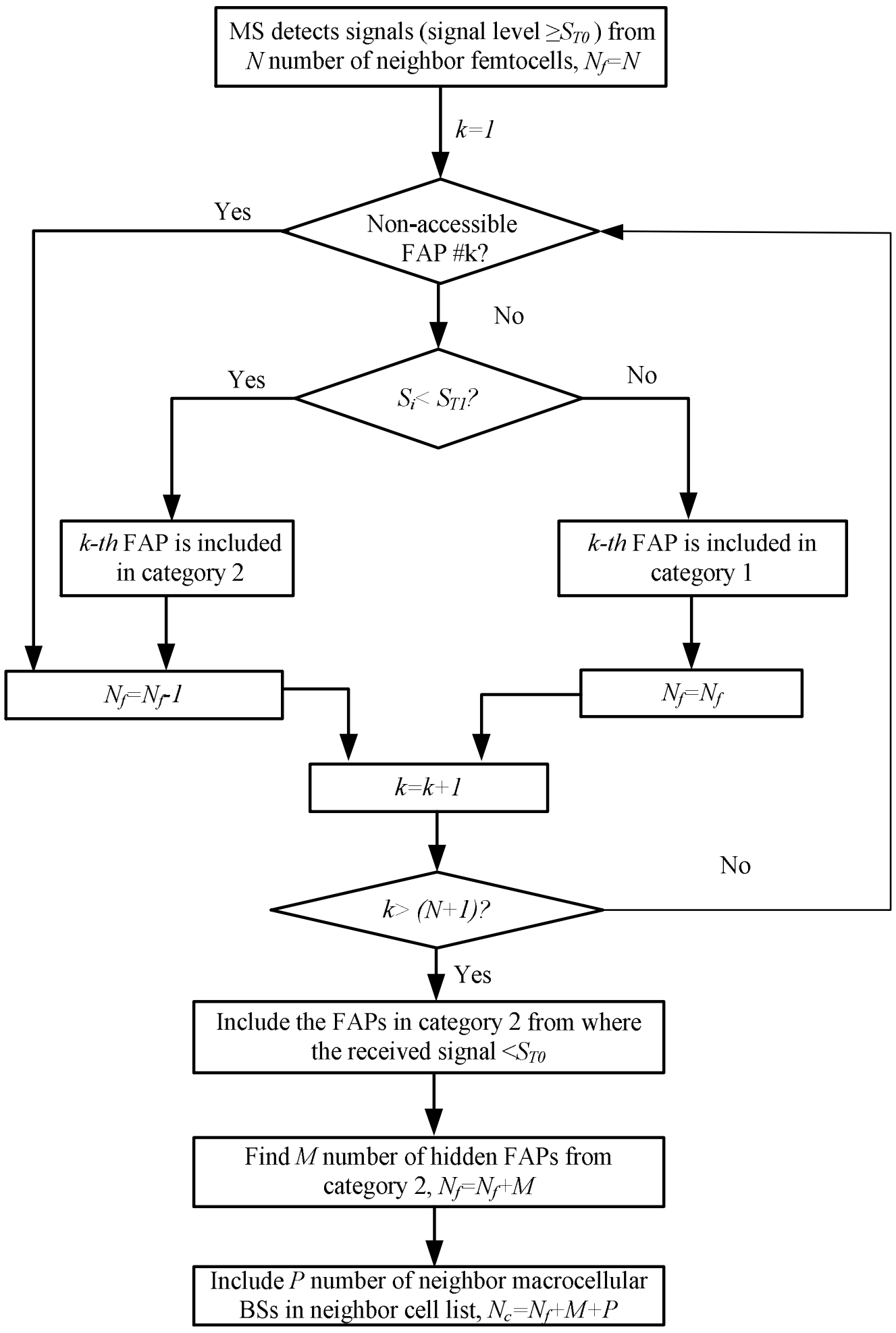

**Fig. 5.** Flow mechanism for the design of the optimal neighbor cell list for handover when the MS is connected with the overlaid macrocellular BS.

The FAPs that are listed in the neighbor femtocell list based only on the received RSSI level can be expressed as set $A$:

$$A = \{...FAP\#i(RSSI_i),....: 1 \leq i,\ RSSI_i \geq S_{T0}\}. \tag{1}$$

where $FAP\#i(RSSI_i)$ represents that *i-th* neighbor FAP from which the received RSSI level at the MS is greater than or equal to $S_{T0}$. $S_{T0}$ is the minimum level of the received signal from an FAP that can be detected by an MS.

The number of FAPs listed based only on the minimum level of received signal level, $S_{T0}$, can be calculated as follows:

$$N = \left|\{...FAP\#i(RSSI_i),....: 1 \leq i,\ RSSI_i \geq S_{T0}\}\right| \quad . \tag{2}$$

Instead of considering only the RSSI level, we consider the RSSI level, frequency used by the serving FAP and *i-th* neighbor FAP, and the location information to construct an appropriate neighbor femtocell list.

In dense femtocells environments, we need to reduce unnecessary handovers. Normally, unnecessary handovers occur owing to the movement of users at the edge of femtocell coverage. We consider a slightly higher RSSI level $S_{T1}$, instead of $S_{T0}$, to reduce unnecessary handovers as well as the ping-pong effect. However, if some FAPs are close to the MS but the signal levels are less than $S_{T1}$ owing to obstacles, these hidden femtocells are picked for the neighbor femtocell list with the coordinated help of the serving FAP and the hidden FAPs. The FAPs with an RSSI level of $S_{T1}$ in the neighbor femtocell list can be expressed as follows:

$$B = \{...FAP\#j(RSSI_j),....: 1 \leq j,\ RSSI_j \geq S_{T1}\}. \tag{3}$$

The number of FAPs listed based on the minimum level of received signal $S_{T1}$ can be calculated as follows:

$$N_1 = \left|\{...FAP\#j(RSSI_j),....: 1 \leq j,\ RSSI_j \geq S_{T1}\}\right|. \tag{4}$$

In dense femtocell deployment, the same frequency is not used for overlapped femtocells [5, 6]. Therefore, for the femtocell-to-femtocell handover case, we can deduct those femtocells from the neighbor femtocell list that use the same frequency as the serving femtocells. The femtocells that can be categorized into this group are

$$C = \{...FAP\#k(f_k),....: 1 \leq k,\ C \in B, f_s \cup f_i = f_s\}, \tag{5}$$

$$N_2 = \left|\{...FAP\#k(f_k),....: 1 \leq k, C \in B, f_s \cup f_i = f_s\}\right|, \tag{6}$$

where $FAP\#k(f_k)$ represents the *k-th* neighbor femtocell that uses frequency $f_k$, whereas $f_s$ is the frequency used by the serving femtocell. $N_2$ denotes the number of femtocells in this group. For the macrocell-to-femtocell handover case, if two or more neighbor femtocells from which the MS receives signals use the same frequency, then the femtocells except the nearest one will be included in this group.

Now, we use the location information for the neighbor femtocell list in order to include hidden FAPs in the neighbor femtocell list. The hidden femtocells are chosen from category-2 femtocells. The included femtocells in this category are (a) the femtocells from which the received RSSI levels are less than $S_{T1}$ or (b) the femtocells that use the same frequency as the serving femtocell. Because the serving FAP can coordinate with some of the nearest FAPs, [6, 13] the nearest FAPs can identify the location of some of the hidden FAPs. Thus, the hidden FAPs within a range of distance can be included in the neighbor femtocell list. The femtocells that are included in this group can be expressed as

$$D=\left\{...FAP\#m\left(RSSI_m,f_m,d_m\right),....:1\le m,\left(\left(RSSI_m<S_{T1}\right)\vee\left(f_s\cup f_m=f_s\right)\right)\wedge\left(d_m\le d_{max}\right)\right\}, \tag{7}$$

$$M=\left|\left\{...FAP\#m\left(RSSI_m,f_m,d_m\right),....:1\le m,\left(\left(RSSI_m<S_{T1}\right)\vee\left(f_s\cup f_m=f_s\right)\right)\wedge\left(d_m\le d_{max}\right)\right\}\right|, \tag{8}$$

where $d_m$ is the distance between the MS and the *m-th* neighbor femtocell that uses frequency $f_m$. The *m-th* femtocell is included in this group only if the distance between the MS and the *m-th* neighbor FAP is less than or equal to a pre-defined threshold distance $d_{max}$.

Considering the above three facts (RSSI level, frequency, and location information), the femtocells included in the final neighbor femtocell list are

$$E=\left(B/C\right)\cup D. \tag{9}$$

The total number of femtocells in the neighbor femtocell list is thus

$$N_f=N_1-N_2+M\ . \tag{10}$$

## 4. Handover Call Flow

To date, an effective and complete handover scheme for femtocell network deployment has been an open research issue. The handover procedures for existing 3GPP networks are presented in [21-27]. In our previous work [28], we presented the handover scheme for small-scale femtocellular network deployment. This section proposes the complete handover call flows for the integrated femtocell/macrocell network architecture in a dense femtocellular network deployment. The proposed handover schemes optimize the selection/reselection/radio resource control (RRC) management functionalities in the femtocell/macrocell handover.

Macrocell-to-femtocell and femtocell-to-femtocell handovers suffer from some additional challenges because each macrocell coverage area may have thousands of femtocells. In these handovers, the MS needs to select the appropriate target FAP among many FAPs. In addition, the interference level should be considered for handover decision. Handover from femtocell-to-macrocell does not have additional complexity as compared with traditional handovers. The basic procedures for handovers in the dense femtocellular network deployment include signal level measurement, SON configuration, optimized neighbor cell list, selection of appropriate access network for the handover, handover decision, and handover execution.

### *4.1 Femtocell-to-Macrocell Handover*

Fig. 6 shows the detailed call flow procedures for femtocell-to-macrocell handover in dense femtocellular network deployment. If a femtocell user detects that the femto signal is going down, the MS sends the report to the connected FAP (steps 1 and 2). The MS searches for the signals from the neighboring FAPs and the macrocellular BS (step 3). The MS, serving FAP (S-FAP), neighbor FAPs, and the macrocellular BS together perform the SON configuration to create an optimized neighbor cell list for the handover (steps 4 and 5). The MS performs pre-authentication with all the access networks that are included in the neighbor cell list (step 6). On the basis of pre-authentication and the received signal levels, the MS and S-FAP together decide to handover to the macrocellular BS (step 7). The FAP starts handover (HO) procedures by sending a handover request to the macrocellular BS through the CN (steps 8–11). CAC and RRC are performed to check whether the call can be accepted or not (step 12). Then, the macrocellular BS responds to the handover request (steps 13–16). Steps 17–21 are used to setup a new link between the target RNC (T-RNC) and the macrocellular BS. The packet data are forwarded to the macrocellular BS (step 22). The MS re-establishes a channel with the macrocellular BS, detaches from the S-FAP, and synchronizes with the macrocellular BS (steps 23–27). The MS sends a handover complete message to the FGW to inform it that the MS has already completed handover and synchronizes with the target macrocellular BS (steps 28–30). Then, the FAP deletes the old link with the S-FAP (steps 31–33). The packets are then sent to the MS through the macrocellular BS.

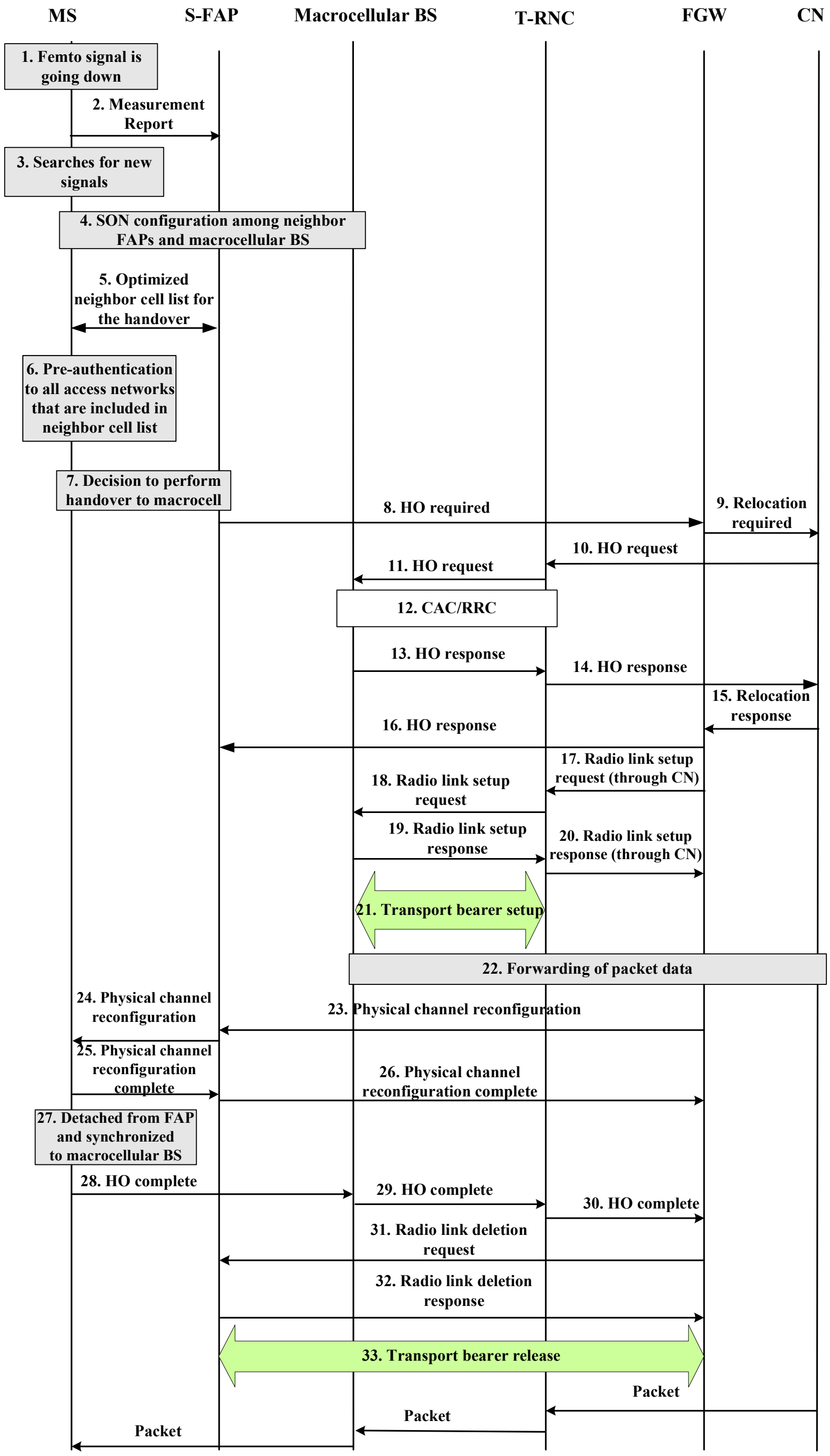

**Fig. 6.** Call flow for the femtocell-to-macrocell handover for a dense femtocellular network deployment.

### *4.2 Macrocell-to-Femtocell Handover*

In this handover, the MS needs to select the appropriate target FAP (T-FAP) among many candidate FAPs. In addition, the interference level should be monitored for handover decision. The authorization should be checked during the handover preparation phase. Fig. 7 details the call flow procedures for macrocell-to-femtocell handover in dense femtocellular network deployment. Whenever the MS in the macrocell network detects a signal from femtocell, it sends a measurement report to the connected macrocellular BS (steps 1 and 2). The combination of the MS, macrocellular BS, and neighbor FAPs perform the SON configuration to create an optimized neighbor cell list for the handover (steps 3 and 4). The MS performs pre-authentication with all the access networks that are included in the neighbor cell list (step 5). On the basis of the pre-authenticated and received signal levels, the MS decides to handover to the T-FAP (step 6).

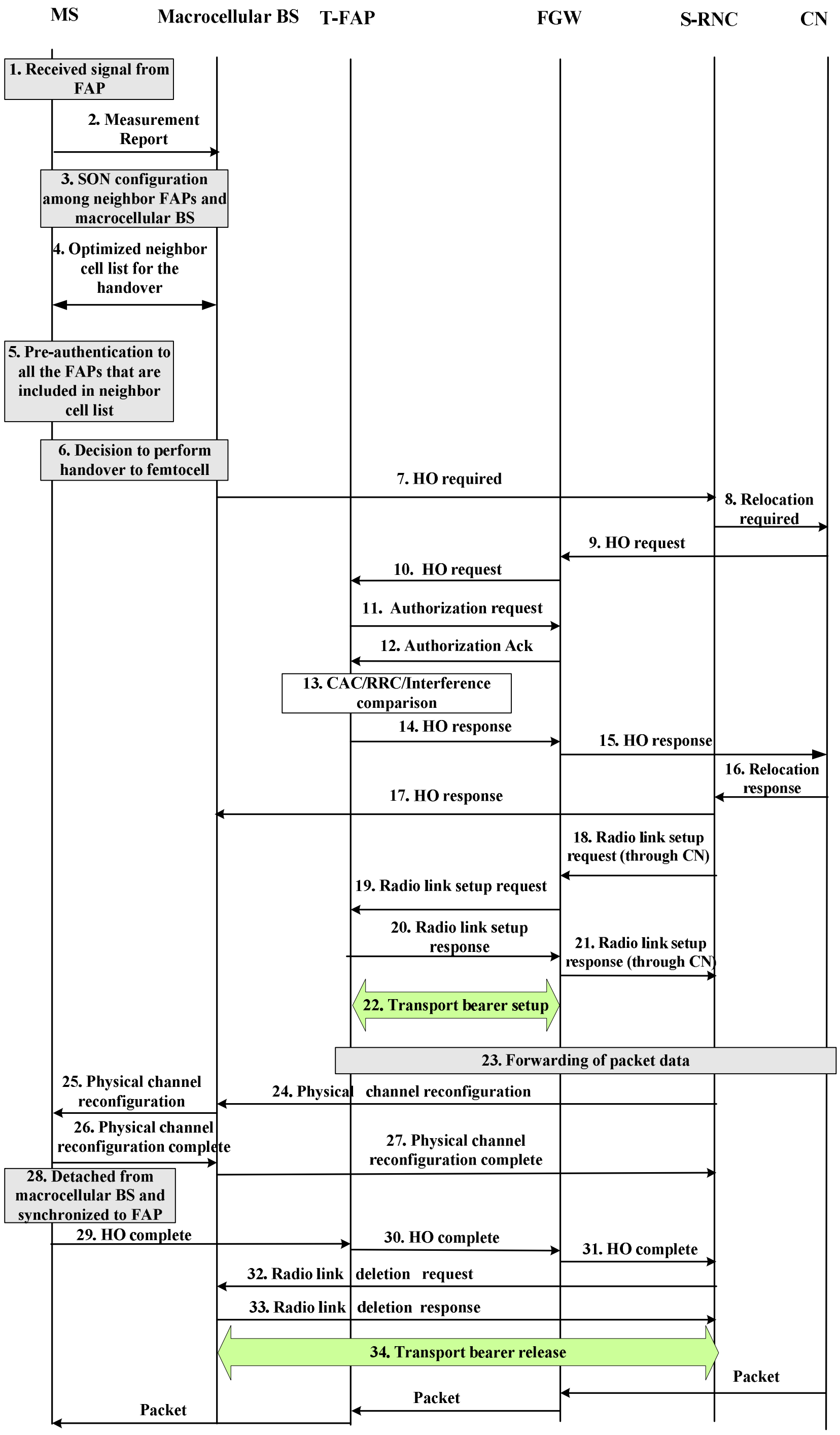

**Fig. 7.** Call flow for the macrocell-to-femtocell handover for dense femtocellular network deployment.

The macrocellular BS starts the handover procedures by sending a handover request to the serving RNC (S-RNC) (step 7). The handover request is forwarded from the macrocellular BS to the T-FAP through the CN and FGW (steps 8–10). The FAP checks the user's authorization (steps 11 and 12). The T-FAP performs CAC, RRC, and compares the interference levels to admit a call (step 13). Then, the T-FAP responds to the handover request to the macrocellular BS through the CN (steps 14–17). A new link is established between the FGW and the T-FAP (steps 18–22). Then, the packet data are forwarded to the T-FAP (step 23). Now, the MS re-establishes a channel with the T-FAP, detaches from the source macrocellular BS, and synchronizes with the T-FAP (steps 24–28). The MS sends a handover complete message to S-RNC to inform it that the MS already completed the handover and synchronized with the T-FAP (steps 29–31). Then, the macrocellular BS deletes the old link with the RNC (steps 32–34). Now, the packets are forwarded to the MS through the FAP.

### *4.3 Femtocell-to-Femtocell Handover*

In this handover, the MS needs to select the appropriate T-FAP among many neighbor FAPs. The authorization should be checked during the handover preparation phase. Fig. 8 shows the detailed call flow procedures for the femtocell-to-femtocell handover in a dense femtocellular network environment. If a femtocell user detects that the femto signal is going down, the MS sends a report to the connected FAP (steps 1 and 2). The MS searches for the signals from the neighbor FAPs and the macrocellular BS (step 3). The MS, S-FAP, neighbor FAPs, and the macrocellular BS perform the SON configuration to create an optimized neighbor cell list for the handover (steps 4 and 5). The MS performs pre-authentication with all the access networks that are included in the neighbor cell list (step 6). On the basis of the pre-authentication and the received signal levels, the MS and S-FAP decide to handover to the T-FAP (step 7). The S-FAP starts handover procedures by sending a handover request to the T-FAP through the FGW (steps 8 and 9). The T-FAP checks the user's authorization (steps 10 and 11). The T-FAP performs CAC and RRC to admit the handover call (steps 12). Then, the T-FAP responds to the handover request from the S-FAP through the FGW (step 13 and 14). A new link is established between the FGW and the T-FAP (steps 15–17). Then, the packet data are forwarded to the T-FAP (step 18). Now, the MS re-establishes a channel with the T-FAP, detaches from the S-FAP, and synchronizes with the T-FAP (steps 19–23). The MS sends a handover complete message to the FGW to inform it that the MS has already completed handover and synchronized with the T-FAP (steps 24–26). Then, the S-FAP deletes the old link with the FGW (steps 27–29). Now, the packets are forwarded to the MS through the T-FAP.

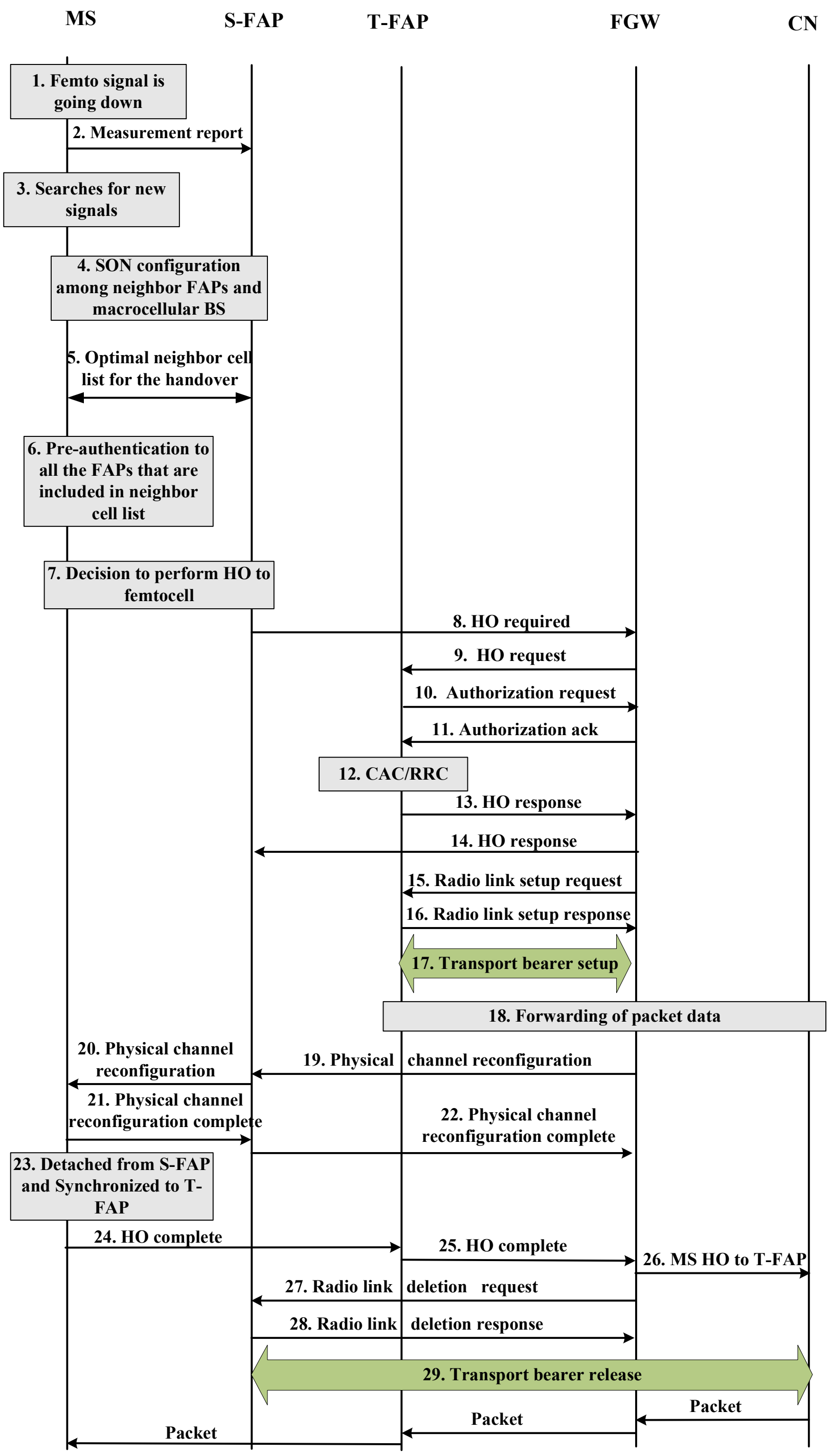

**Fig. 8**. Call flow for the femtocell-to-femtocell handover for dense femtocellular network deployment.

## 5. CAC for Femtocell/Macrocell Overlaid Networks

For the femtocell/macrocell integrated networks, the CAC can play a vital role in maximizing resource utilization, particularly for macrocellular networks, by efficiently controlling the admission of various traffic calls inside the macrocell coverage area. The main objective of our proposed scheme is to transfer a larger number of macrocell calls to femtocellular networks. We divide the proposed CAC into three parts. The first one is for the new originating calls, the second one is for the calls that are originally connected with the macrocellular BS, and the third one is for the calls that are originally connected with the FAPs. We also use two threshold levels of SNIR to admit a call in the system. The first threshold level $\Gamma_1$ is the minimum level of the received SNIR that is needed to connect a call to any FAP. The second signal level $\Gamma_2$ is higher than $\Gamma_1$. The second threshold is used in the CAC to reduce the unnecessary macrocell-to-femtocell handovers. We offer QoS degradation [14, 15] of the QoS adaptive multimedia traffic to accommodate femtocell-to-macrocell and macrocell-to-macrocell handover calls. The existing QoS adaptive multimedia traffic in overlaid macrocellular network releases $C_{release}$ amount of bandwidth to accept the handover calls in the macrocellular network. This releasable amount depends on the number of running QoS adaptive multimedia calls and their maximum level of allowable QoS degradation and the total number of existing calls in the macrocellular network. Suppose $\beta_{r,m}$ and $\beta_{min,m}$ are the requested bandwidth by a call and the minimum allocated bandwidth for a call of traffic class *m*, respectively. Then, each of the *m-th* class QoS adaptive calls can release a maximum ($\beta_{r,m}$ - $\beta_{min,m}$) amount of bandwidth to accept a call in the macrocell system. If $C$ and $C_{occupied}$ are the macrocell system bandwidth capacity and the occupied bandwidth by the existing macrocell calls, respectively, then the available empty bandwidth $C_{available}$ in the macrocellular network is ($C$ - $C_{occupied,m}$).

### *5.1 New Originating Calls*

Fig. 9 shows the CAC policy for new originating calls. Whenever a new call arrives, the CAC initially checks whether the femtocell coverage is available or not. If femtocell coverage is available, then an FAP is the first choice to connect a call. An FAP accepts a new originating call if the received SNIR level $\Gamma_2$ is satisfied and resources in the FAP are available. $SNIR_{T,f}$ is the received SNIR level of the target FAP. If the above conditions are not satisfied, then the call tries to connect with the overlaid macrocellular network. The macrocell system does not allow the QoS degradation policy to accept any new originating calls. A call of *m-th* class traffic is rejected if the requested bandwidth $\beta_{r,m}$ is not available in the overlaid macrocellular network.

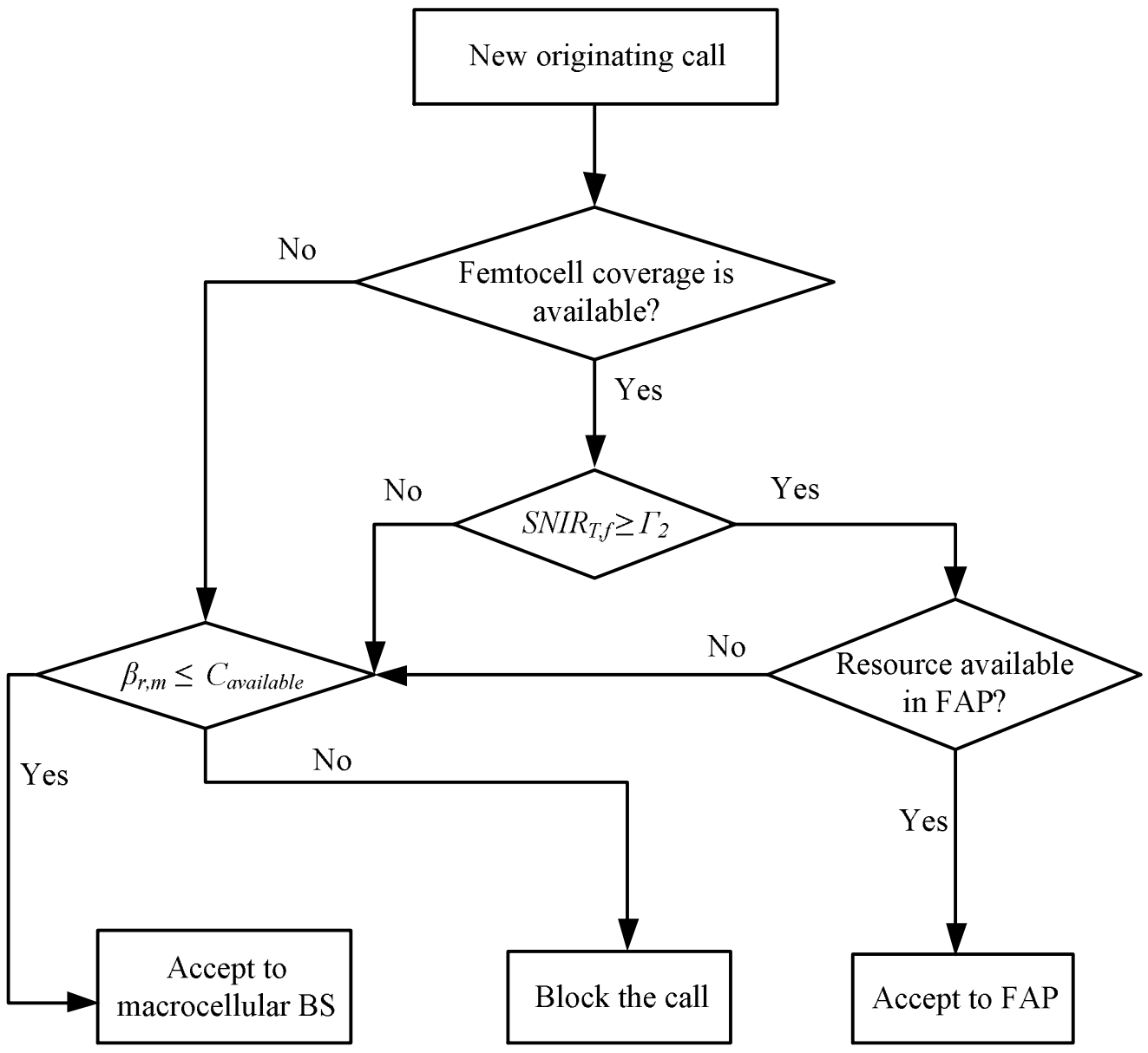

**Fig. 9.** CAC policy for new originating calls.

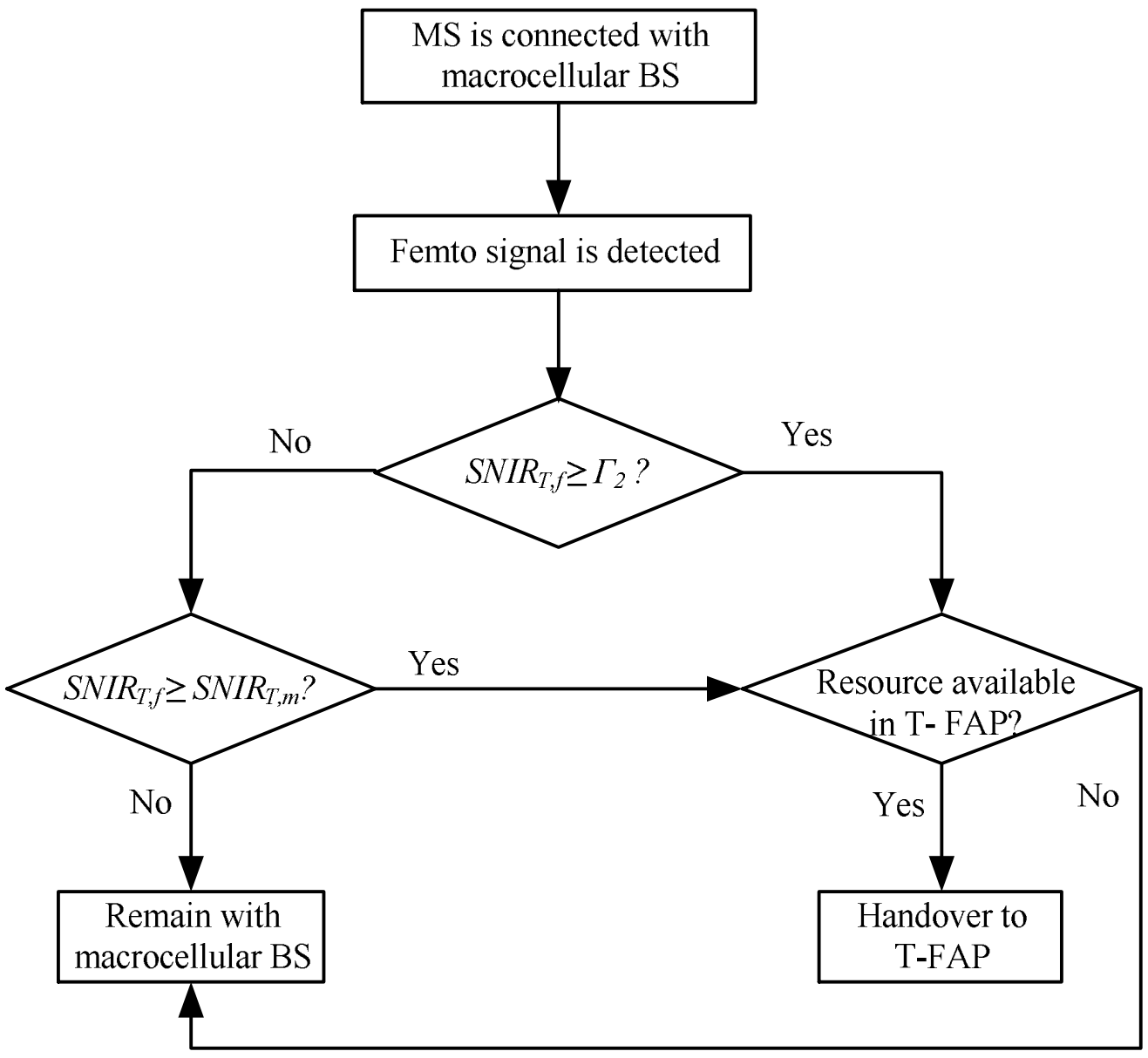

**Fig. 10.** CAC policy for the calls that are originally connected with the macrocellular BS.

## *5.2 Calls that are Originally Connected with the Macrocellular BS*

Fig. 10 shows the CAC policy for the calls that are originally connected with the macrocellular BS. Whenever the moving MS detects a signal from an FAP, the CAC policy checks the received SNIR level, i.e., $SNIR_{T,f}$, for the target FAP. A macrocell call is handed over to the femtocell if the $SNIR_{T,f}$ meets the minimum $\Gamma_2$ or the currently received SNIR level of the macrocellular BS, $SNIR_m$, is less than or equal to $SNIR_{T,f}$. If any one of the above conditions is satisfied, then the CAC policy checks the resource availability in the target FAP. We prefer the higher level of threshold $\Gamma_2$ to avoid some unnecessary macrocell-to-femtocell handovers.

### *5.3 Calls that are Originally Connected with the FAPs*

Fig. 11 shows the CAC policy for calls that are originally connected with the FAPs. Femtocell-to-femtocell and femtocell-to-macrocell handover calls are controlled by this CAC policy. Whenever the signal level from the S-FAP is going down, the MS initiates a handover to other femtocells or an overlaid macrocell. Whenever another T-FAP is not available for handover, the call tries to connect with the macrocellular network. If an empty resource in the macrocell system is not enough to accept the call, the CAC policy allows the release of some bandwidth from the existing calls by degrading their QoS level. The CAC policy also permits the reduction of the required bandwidth for a handover call request. The system allows a maximum ($\beta_{r,m}$ - $\beta_{min,m}$) amount of bandwidth reduction for an existing call or a requested handover call. Therefore, the system increases the number of calls admitted as well as reduces the handover call dropping probability. If the minimum required bandwidth $\beta_{min,m}$ is not available in the macrocell system after releasing of some bandwidth from the existing calls, then the call is dropped. If the received SNIR of the T-FAP is greater than or equal to $\Gamma_2$, the MS first tries to handover to the T-FAP. Conversely, if the received SNIR of the T-FAP is in between $\Gamma_1$ and $\Gamma_2$, then the MS initially tries to connect with the macrocellular BS. If resources are not available in the macrocell system, the MS attempts to hand over to the T-FAP, even if the received SNIR of the T-FAP is less than $\Gamma_2$. However, during this condition, the QoS degradation policy is not applicable. The QoS degradation policy is only applicable when the received SNIR of the T-FAP is less than $\Gamma_1$ or resources in the T-FAP are not available.

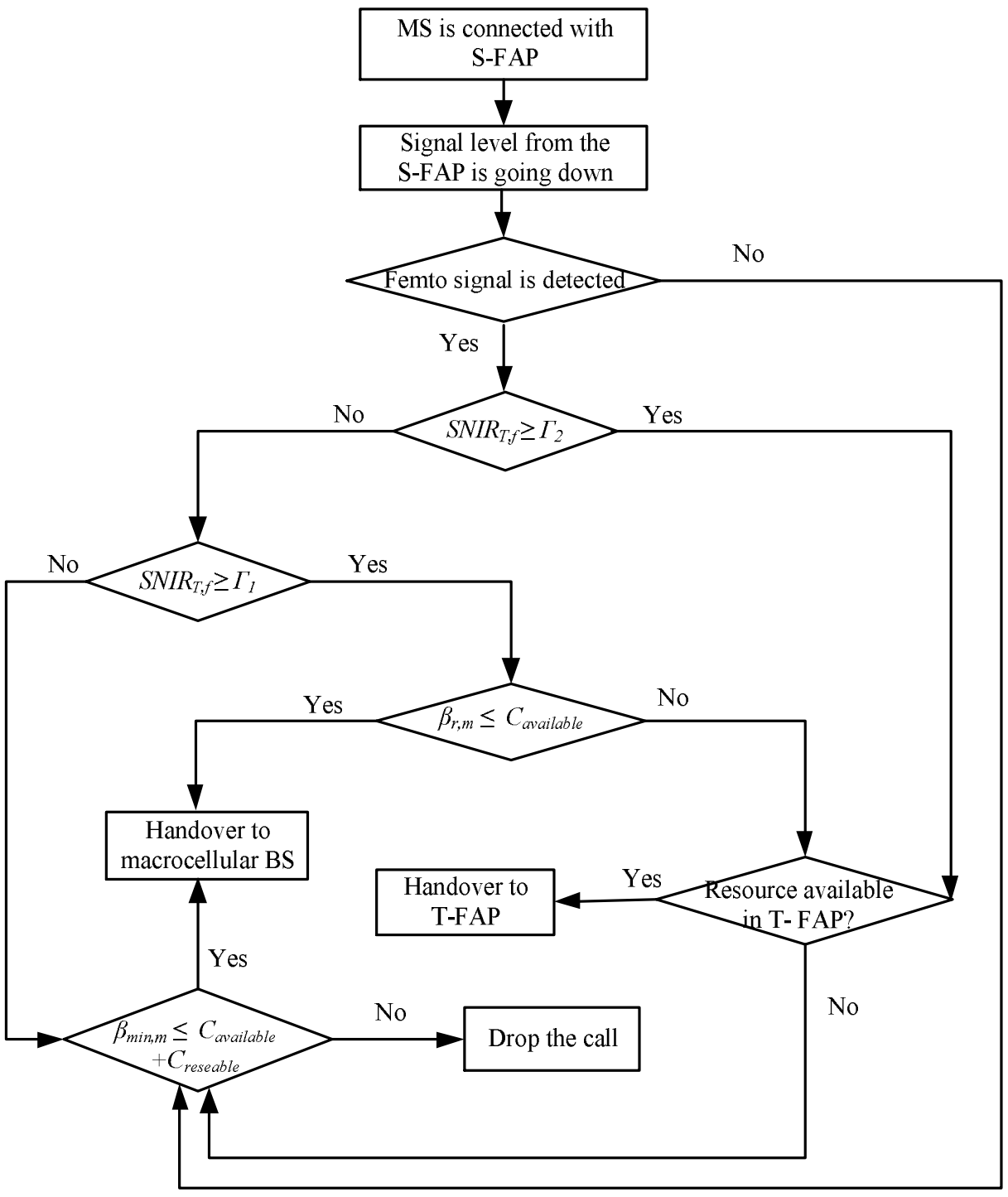

**Fig. 11.** CAC policy for calls that are originally connected with the FAP.

## 6. Queuing Analysis and Traffic Model

The proposed CAC schemes can be modeled by Markov chain. The Markov chain for the queuing analysis of a femtocell layer is shown in Fig. 12, where the states of the system represent the number of calls in the system. The maximum number of calls that can be accommodated in a femtocell system is $K$. As the call arrival rate in a femtocell is normally very low and the data rate of a femtocellular network is high, there is no need for a handover priority scheme for the femtocellular networks. The calls that have arrived in a femtocellular network are new originating calls, macrocell-to-femtocell handover calls, and femtocell-to-femtocell handover calls. Femtocell-to-femtocell handover calls are divided into two types. The first type of call is when the received SNIR of the T-FAP is greater than or equal to $\Gamma_2$. The second type of call is when the received SNIR of the T-FAP is between $\Gamma_1$ and $\Gamma_2$, and these calls are rejected by the macrocellular BS. We define $\mu_m$ ($\mu_f$) as the channel release rate of the macrocell (femtocell).

Fig. 13 shows the Markov chain for the queuing analysis of the overlaid macrocell layer, where the states of the system represent the number of calls in the system. In Figs. 12 and 13, symbols $\lambda_{o,f}$ and $\lambda_{o,m}$ denote the total originating call arrival rates considering all $n$ number of femtocells within a macrocell coverage area and only the macrocell coverage area, respectively. $\lambda_{h,mm}$, $\lambda_{h,ff}$, $\lambda_{h,fm}$, and $\lambda_{h,mf}$ denote the total macrocell-to-macrocell, femtocell-to-femtocell, femtocell-to-macrocell, and macrocell-to-femtocell handover call arrival rates within the macrocell coverage area, respectively. $P_{B,m}$ ($P_{B,f}$) is the new originating call blocking probability in the macrocell (femtocell) system. $P_{D,m}$ ($P_{D,f}$) is the handover call dropping probability in the macrocell (femtocell) system. We assume that for a femtocell-to-femtocell handover, the probability that the received SNIR of the T-FAP is greater $\Gamma_2$ and is represented by $\alpha$, and the received SNIR of the T-FAP is between $\Gamma_2$ and $\Gamma_2$ and is represented by $\beta$.

Fig. 13 also shows that the macrocell system provides $S$ number of additional states to support handover calls by the proposed adaptive QoS policy. State $N$ is the maximum number of calls that can be accommodated by the macrocell system without a QoS adaptation policy. Hence, the system provides a QoS adaptation policy only to accept handover calls in the macrocell system. These handover calls include macrocell-to-macrocell and femtocell-to-macrocell handover calls. Femtocell-to-macrocell handover calls are divided into two types. The first type of call is for those that have directly arrived to the macrocell system. The second type of call is those for which the calls have first arrived to femtocells, but are not accepted to the femtocells owing to lagging of resources or poor SNIR level.

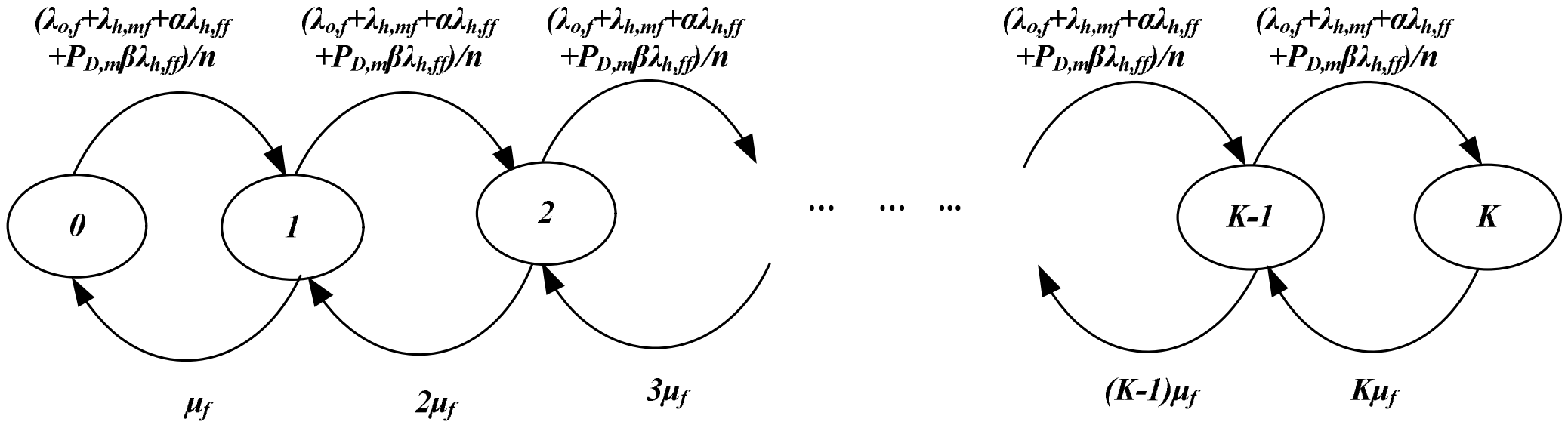

**Fig. 12.** Markov chain of a femtocell layer.

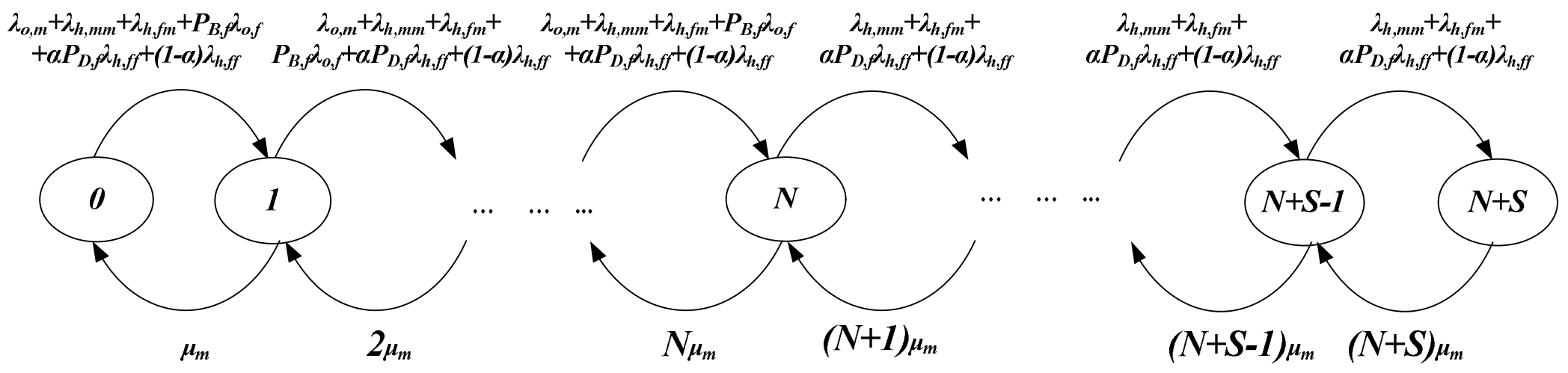

**Fig. 13.** Markov chain of a macrocell layer.

The average channel release rate for the macrocell layer increases as the number of deployed femtocells increases. Because of the increasing number of femtocells, more macrocell users are handed over to femtocell networks. The average channel release rates [29] for the femtocell layer and the macrocell layer are calculated as follows.

For the macrocell layer, the average channel release rate is

$$\mu_m = \eta_m(\sqrt{n}+1)+\mu\,, \tag{11}$$

and for the femtocell layer, it is

$$\mu_f = \eta_f + \mu , \tag{12}$$

where $1/\mu$, $1/\eta_m$, and $1/\eta_f$ are the average call duration (exponentially distributed), average cell dwell time for the macrocell (exponentially distributed), and the average cell dwell time for the femtocell (exponentially distributed), respectively.

Equating the net rate of calls entering a cell and requiring handover to those leaving the cell, the handover call arrival rates are calculated as follows [29].
The macrocell-to-macrocell handover call arrival rate is

$$\lambda_{h,mm} = P_{h,mm} \frac{\left(1-P_{B,m}\right)\left(\lambda_{m,o}+\lambda_{f,o}P_{B,f}\right)+\left(1-P_{D,m}\right)\left\{\lambda_{h,fm}+\lambda_{h,ff}\left(1-\alpha+\alpha P_{D,f}\right)\right\}}{1-P_{h,mm}\left(1-P_{D,m}\right)}, \tag{13}$$

the macrocell-to-femtocell handover call arrival rate is

$$\lambda_{h,mf} = P_{h,mf} \frac{\left(1-P_{B,m}\right)\left(\lambda_{m,o}+\lambda_{f,o}P_{B,f}\right)+\left(1-P_{D,m}\right)\left\{\lambda_{h,fm}+\lambda_{h,ff}\left(1-\alpha+\alpha P_{D,f}\right)\right\}}{1-P_{h,mm}\left(1-P_{D,m}\right)}, \tag{14}$$

the femtocell-to-femtocell handover call arrival rate is

$$\lambda_{h,ff} = P_{h,ff} \frac{\lambda_{f,o}\left(1-P_{B,f}\right)+\lambda_{h,mf}\left(1-P_{D,f}\right)}{1-P_{h,ff}\left(1-P_{D,f}\right)\left\{\alpha+\left(1-\alpha\right)P_{D,m}\right\}}, \tag{15}$$

and the femtocell-to-macrocell handover call arrival rate is

$$\lambda_{h,fm} = P_{h,fm} \frac{\lambda_{f,o}\left(1-P_{B,f}\right)+\lambda_{h,mf}\left(1-P_{D,f}\right)}{1-P_{h,ff}\left(1-P_{D,f}\right)\left\{\alpha+\left(1-\alpha\right)P_{D,m}\right\}}, \tag{16}$$

where $P_{h,mm}$, $P_{h,mf}$, $P_{h,ff}$, and $P_{h,fm}$ are the macrocell-to-macrocell handover probability, macrocell-to-femtocell handover probability, femtocell-to-femtocell handover probability, and femtocell-to-macrocell handover probability, respectively.

The probability of handover depends on several factors such as the average call duration, cell size, and average user velocity. The handover probabilities from a femtocell and to a femtocell in integrated femtocell/macrocell networks also depend on the density of femtocells and the average size of femtocell coverage areas. Hence, on the basis of the basic derivation for handover probability calculations in [29], we derive the formulas for $P_{h,mm}$, $P_{h,mf}$, $P_{h,ff}$, and $P_{h,fm}$ as follows:

$$P_{h,mm} = \frac{\eta_m}{\eta_m + \mu}, \tag{17}$$

$$P_{h,fm} = \left[1 - n\left(\frac{r_f}{r_m}\right)^2\right]\frac{\eta_f}{\eta_f + \mu}, \tag{18}$$

$$P_{h,ff} = (n-1)\left(\frac{r_f}{r_m}\right)^2\frac{\eta_f}{\eta_f + \mu}, \tag{19}$$

$$P_{h,mf} = n\left(\frac{r_f}{r_m}\right)^2\frac{\eta_m\sqrt{n}}{\eta_m\sqrt{n} + \mu}. \tag{20}$$

There is no guard channel for the handover calls in the femtocell layer in our proposed scheme. For the femtocell layer, the average call blocking probability $P_{B,f}$ and the average call dropping probability $P_{D,f}$ can be calculated as [30]

$$P_{D,f} = P_{B,f} = P_f(K) = \frac{\left(\frac{\lambda_{T,f}}{n}\right)^K \frac{1}{K!\mu_f^K}}{\sum_{i=0}^{K}\left(\frac{\lambda_{T,f}}{n}\right)^i \frac{1}{i!\mu_f^i}}, \tag{21}$$

where $\lambda_{T,f} = \lambda_{f,o} + \lambda_{h,mf} + \alpha\lambda_{h,ff} + P_{D,m}\beta\lambda_{h,ff}$.

A QoS adaptation/degradation policy is allowed for the handover calls of a macrocell layer in our proposed scheme. For the macrocell layer, the average call blocking probability $P_{B,m}$ and the average call dropping probability $P_{D,m}$ can be calculated as [30]

$$P_{B,m} = \sum_{i=N}^{N+S} P(i) = \sum_{i=N}^{N+S}\frac{\left(\lambda_{m,0} + \lambda_{h,m}\right)^N\left(\lambda_{h,m}\right)^{i-N}}{i!\mu_m^i}P(0), \tag{22}$$

$$P_{D,m} = P(N+S) = \frac{\left(\lambda_{m,0} + \lambda_{h,m}\right)^N \lambda_{h,m}^S}{(N+S)!\mu_m^{N+S}}P(0), \tag{23}$$

where $\lambda_{h,m} = \lambda_{h,mm} + \lambda_{h,fm} + \alpha P_{D,f}\lambda_{h,ff} + (1-\alpha)\lambda_{h,ff}$

and $P(0) = \left[\sum_{i=0}^{N}\frac{\left(\lambda_{m,0} + \lambda_{m,h}\right)^i}{i!\mu_m^i} + \sum_{i=N+1}^{N+S}\frac{\left(\lambda_{m,0} + \lambda_{m,h}\right)^N\left(\lambda_{m,h}\right)^{i-N}}{i!\mu_m^i}\right]^{-1}$.

## 7. Performance Analysis

In this section, we studied the effect of integrated femtocell/macrocell networks as well as the performance analysis of our proposed schemes. All the call arriving processes are assumed to be Poisson. The positions of the deployed femtocells within the macrocell coverage area are random. Table 1 lists the basic parameters that are used for performance analysis. We also assume a random distribution of hidden femtocells. We consider both open access and closed access randomly in the simulation. The propagation models used for the analysis are as follows.

The propagation model for the femtocell [31] is

$$L_{femto} = 20\log_{10} f + N\log_{10} d + L_f(n) - 28 \;\; dB \; . \tag{24}$$

The propagation model for the macrocell [32] is

$$\begin{aligned} L_{macro} &= 36.55 + 26.16\log_{10} f - 3.82\log_{10} h_b - a(h_m) + [44.9 - 6.55\log_{10} h_b]\log_{10} d \\ &\quad + L_{sh} + L_{pen} \;\; dB \end{aligned} \; . \tag{25}$$

**Table 1.** Summary of the parameter values used in our analysis

| **Parameter** | **Value** |
|---|---|
| Radius of femtocell coverage area | 10 (m) |
| Carrier frequency for femtocells | 1.8 (GHz) |
| Transmit signal power by macrocellular BS | 1.5 kW |
| Maximum transmit power by an FAP | 10 (mW) |
| Height of macrocellular BS | 100 (m) |
| Height of an FAP | 2 (m) |
| Height of an MS | 2 (m) |
| First threshold value of received signal (RSSI) from an FAP *($S_{T0}$)* | -90 (dBm) |
| Second threshold value of received signal (RSSI) from an FAP *($S_{T1}$)* | -75 (dBm) |
| Bandwidth capacity of a macrocell (C) | 6 (Mbps) |
| Required/allocated bandwidth for each of the QoS non-adaptive calls | 64 (Kbps) |
| Maximum required/allocated bandwidth for each of the QoS adaptive calls | 56 (Kbps) |
| Minimum required/allocated bandwidth for each of the QoS adaptive calls | 28 (Kbps) |
| Ratio of traffic arrivals (QoS non-adaptive calls: QoS adaptive calls) | 1:1 |
| First SNIR threshold ($\Gamma_1$) | 10 (dB) |
| Second SNIR threshold ($\Gamma_2$) | 12 (dB) |
| Number of deployed femtocells in a macrocell coverage area | 1000 |
| Average call duration time ($1/\mu$) considering all calls (exponentially distributed) | 120 (s) |
| Average cell dwell time ($1/\eta_f$) for the femtocell (exponentially distributed) | 360 (s) |
| Average cell dwell time ($1/\eta_m$) for the macrocell (exponentially distributed) | 240 (s) |
| Density of call arrival rate (at femtocell coverage area:at macrocell only coverage area ) | 20:1 |
| Standard deviation for the lognormal shadowing loss | 8 (dB) |
| Penetration loss | 20 (dB) |

First, we compare the performance of the proposed neighbor cell list management scheme. We consider traditional schemes (e.g., [18, 19]) to compare to the performance of our proposed scheme. We assume that the "traditional scheme" includes an FAP or a macrocellular BS in the neighbor cell list if the received signal level from that FAP or macrocellular BS is greater than or equal to $s_{T0}$. Fig. 14 shows the probability comparison that the target femtocell is missing from the neighbor femtocell list for the femtocell-to-femtocell handover. A traditional neighbor cell list cannot include the hidden femtocells in the neighbor cell list based only on the received signal strength. Thus, there is a possibility that the target femtocell is not included in the neighbor femtocell list. This causes a failure of the handover to the target femtocell. Increasing the number of deployed femtocells within an area increases the possibility that the neighboring FAPs coordinate with the serving FAP and stay informed of the location of the hidden neighbor femtocells. As a consequence, an increased number of deployed femtocells results in the reduction of probability that the hidden femtocells are out of the neighbor femtocell list. Moreover, missing the appropriate neighbor femtocell from the neighbor femtocell list may cause a handover failure. Thus, the handover failure rate decreases with an increase in the number of deployed femtocells in the proposed scheme. Fig. 15 shows the comparison of the numbers of neighbor femtocells in the neighbor femtocell list for the femtocell-to-femtocell handover. The result shows that the neighbor femtocell list based on the proposed scheme contains a very small number of femtocells during the handovers. Thus, the number of signal flows for the handover process becomes very small. Therefore, the results in Figs. 14 and 15 show that the proposed neighbor femtocell list algorithms for the femtocell-to-femtocell and the macrocell-to-femtocell handovers offer an optimal number of femtocells in the neighbor femtocell list. However, the reduced number of femtocells in the neighbor femtocell list does not increase the handover failure probability.

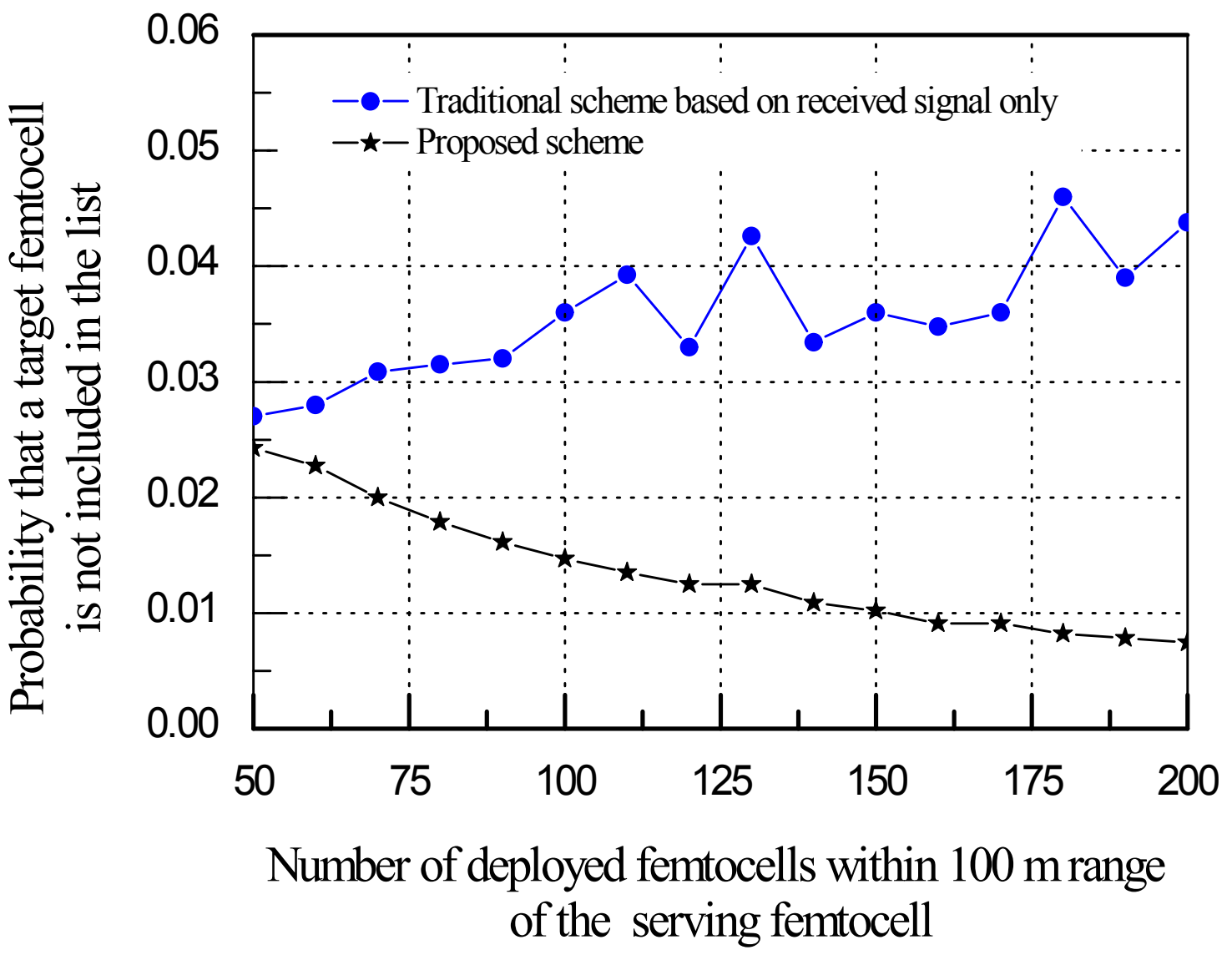

**Fig. 14.** Probability comparison that the target femtocell is missing from the neighbor femtocell list (considering the femtocell-to-femtocell handover).

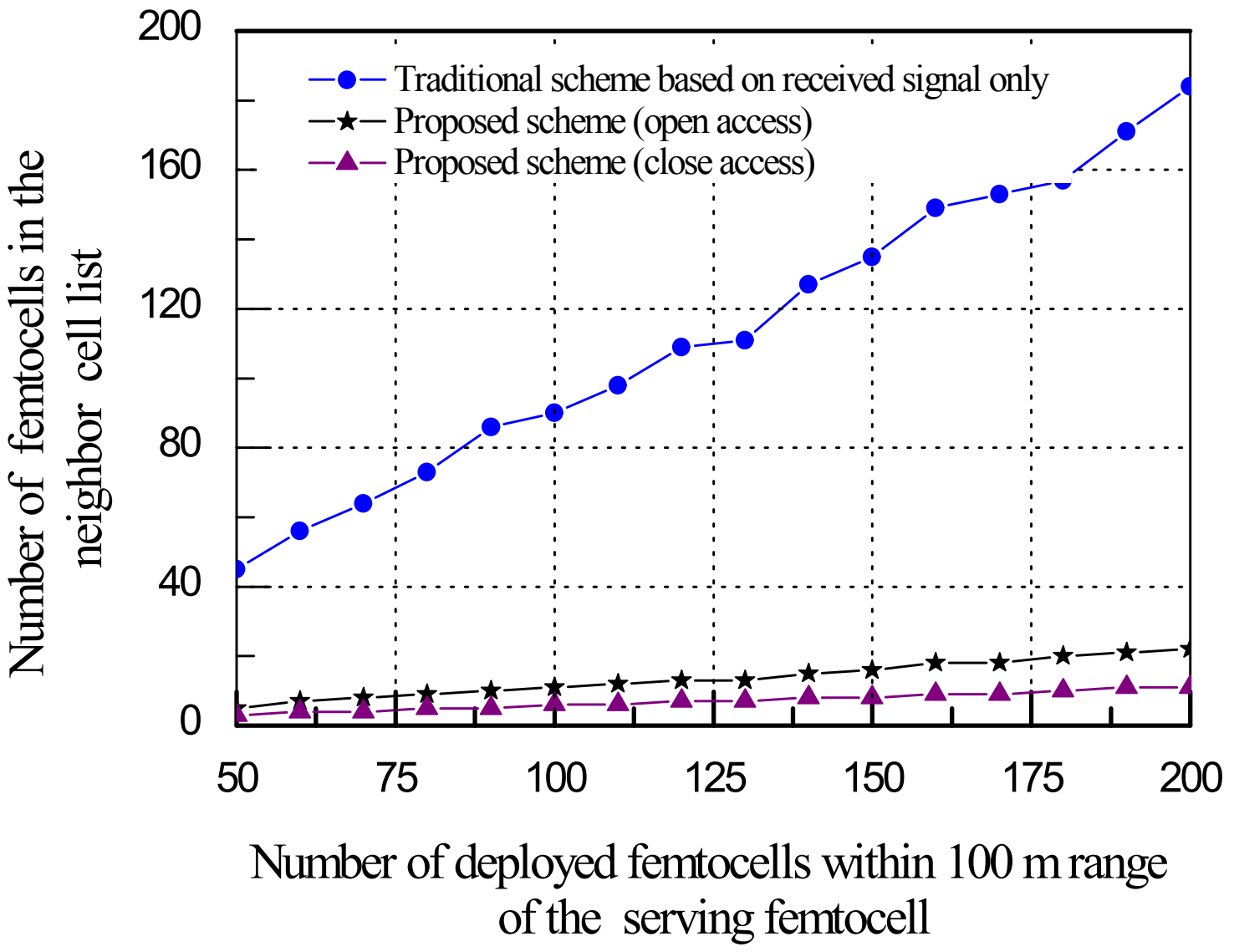

**Fig. 15.** Comparison of the number of neighbor femtocells in the neighbor femtocell list for different schemes based on different parameters metrics (considering the femtocell-to-femtocell handover).

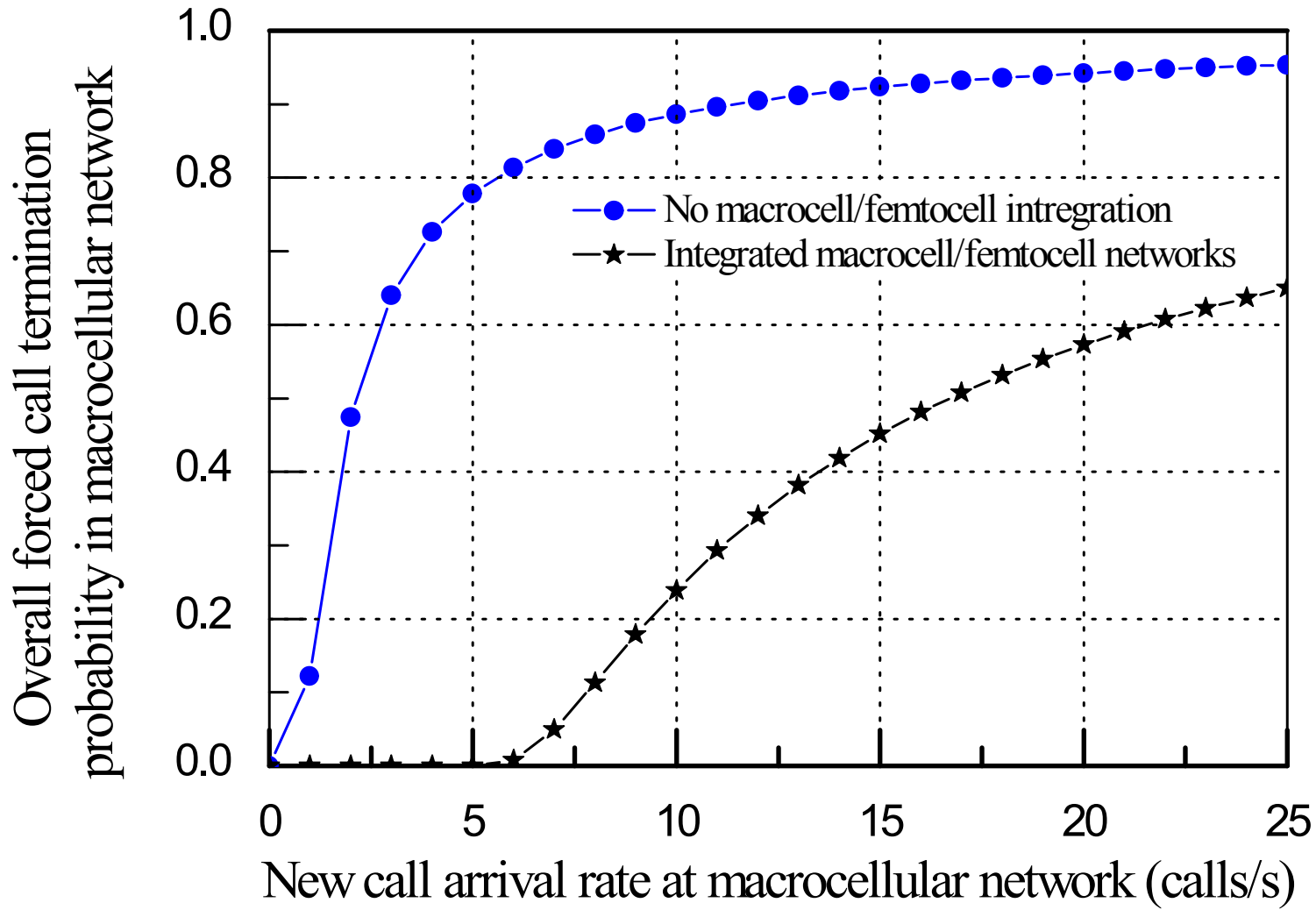

**Fig. 16.** Comparison of overall forced call termination probability in the macrocell system.

Whenever the macrocell and the femtocells are integrated, a large number of macrocell calls are diverted to femtocells through the macrocell-to-femtocell handover. As a result, the macrocell system can accommodate a larger number of calls. Fig. 16 shows the performance improvement of macrocellular networks in terms of the overall forced call termination probability. Fig. 17 shows the effect of different handover probabilities with an increase in the number of deployed femtocells within a macrocellular network coverage. With an increase in the number of deployed femtocells, the femtocell-to-femtocell handover and macrocell-to-femtocell handover probabilities are significantly increased. In addition, the femtocell-to-macrocell handover probability is very high. Thus, the management of these large number of handover calls is the important issue for dense femtocellular network deployment. Fig. 18 shows the effect of the femtocell/macrocell integrated networks in terms of the channel release rate of the macrocellular network. Owing to the integration, a large number of macrocell users are handed over to femtocellular networks. Thus, the channel release rate increases with an increase in the number of deployed femtocells. As a consequence, the macrocellular network can significantly reduce the overall forced call termination probability.

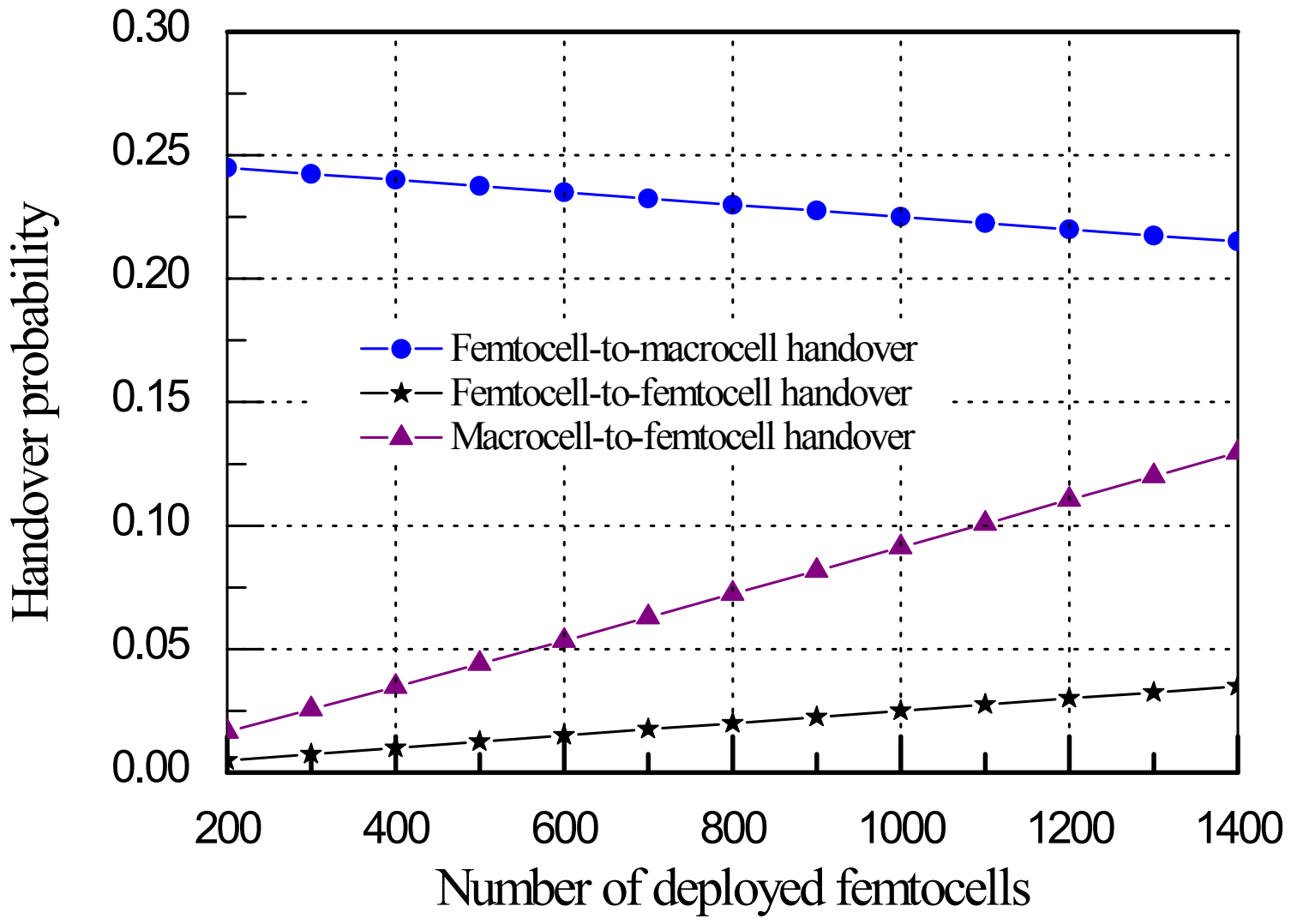

**Fig. 17.** Comparison of handover probability.

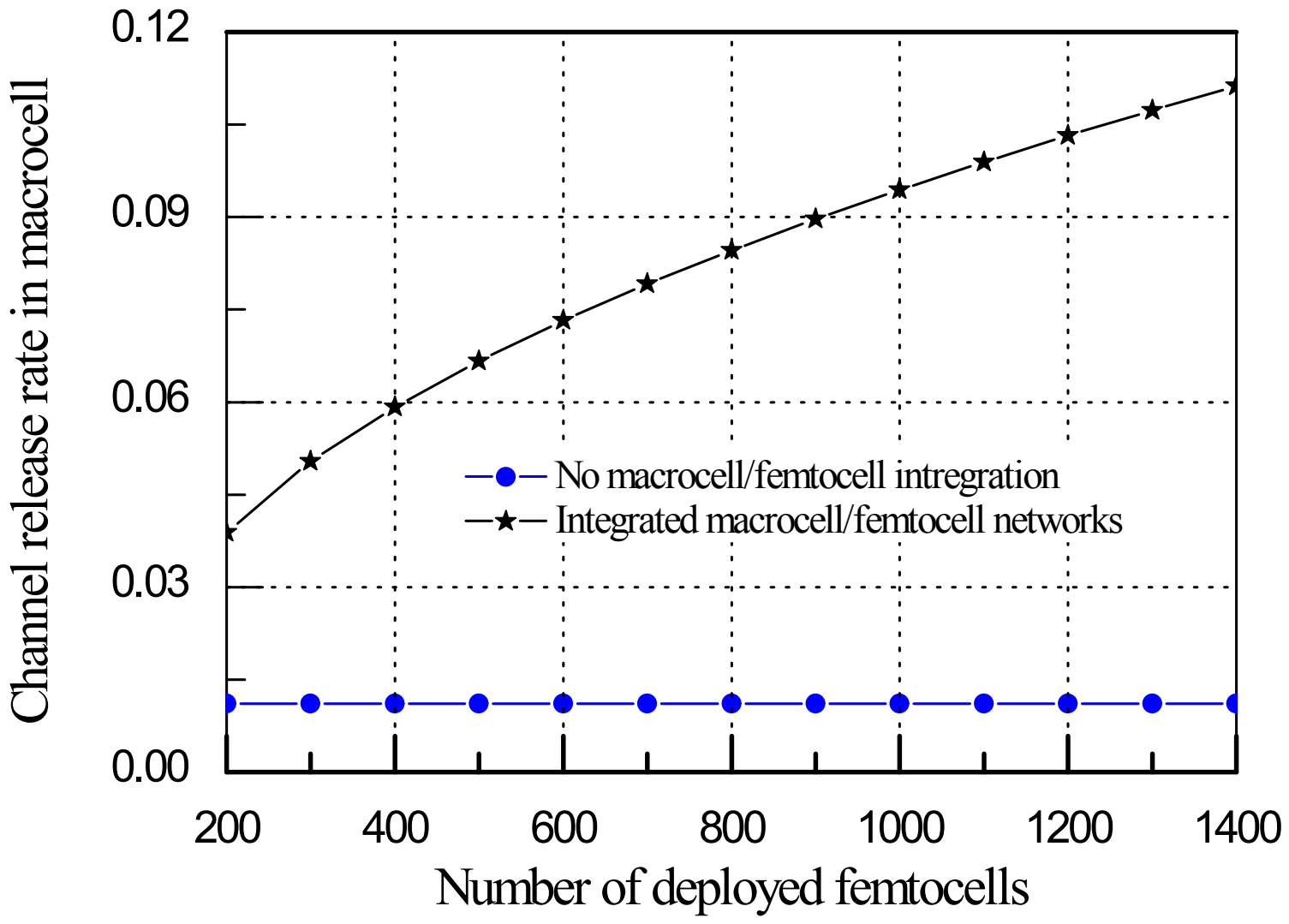

**Fig. 18.** Comparison of channel release rate in an overlaid macrocellular network.

The results in Fig. 14–Fig. 18 show the improvement of the proposed schemes. Our proposed neighbor cell list algorithms provide an efficient way to manage the neighbor cell list. The reduced number of FAPs in the neighbor cell list results in reduced scanning and signaling. The inclusion of hidden FAPs in the neighbor cell list results in reduced handover failure probability to the femtocell. The proposed QoS adaptive/degradation policy is able to handle a large number of handover calls. The integration of a macrocell with the femtocells provides reduced overall forced call termination probability in the macrocell system. The integrated femtocell/macrocell network system also increases the macrocell channel release rate that results in an increased load transfer rate from the macrocellular network to the femtocellular networks.

## 8. Conclusion and Future Research

Femtocellular networks may have different sizes, and ultimately, we expect to see densely deployed networks with over thousands of femtocells overlaid by a single macrocell. Mobility management is one of the key issues for successful dense femtocellular network deployment. However, a complete solution for the mobility management for femtocellular networks is still an open research issue. We proposed novel approaches to solve the mobility management issues for densely deployed femtocellular networks. The proposed SON-based network architecture is capable of handling large numbers of FAPs inside the macrocell coverage. Our proposed algorithm helps overcome the hidden FAP problem. The reduced neighbor cell list results in reduced power loss as well as reduced MAC overhead. The proposed handover call flows will be very effective to implement for handover processes in dense femtocellular network deployment. The suggested traffic model for the femtocell/macrocell integrated network is quite different from the existing macrocellular network traffic model. This traffic model can be applied for the performance analysis of a femtocell/macrocell integrated network. The results shown in this paper clearly imply the advantages of our proposed schemes. The analyses also indicate the effect of femtocellular network deployment and performance improvement attributed to the integrated femtocell/macrocell network. Therefore, our performance analyses show that mobility management is a critical issue for dense femtocellular network deployment.

We studied major research issues concerning mobility management in integrated femtocellular/macrocellular networks. The research results were studied using several numerical and simulation analyses. A real-life experiment would require many FAPs as testing equipment. Therefore, experimental results for comparison to theory are saved for future research work. However, our proposed scheme provides a good basis for research as well as industry to implement dense femtocells successfully.

**Acknowledgements** This work was supported by the IT R&D program of MKE/KEIT. [10035362, Development of Home Network Technology based on LED-ID].